\documentclass[fleqn,usenatbib,useAMS]{mnras}
\pdfoutput=1		

\usepackage{txfonts}

\usepackage[T1]{fontenc}
\usepackage{ae,aecompl}

\usepackage{amsmath}
\usepackage{graphicx}
\usepackage{amssymb}
\usepackage{xspace}
\usepackage{natbib}
\usepackage{dcolumn}

\newcolumntype{d}{D{=}{-}{-1}}		
\newcolumntype{p}{D{+}{\pm}{-1}}		
\newcommand{\dev}{de Vaucouleurs\xspace}		
\newcommand{\ser}{S\'{e}rsic\xspace}		
\newcommand{\tld}{\ifmmode\textstyle{\sim}\else$\sim$\fi}	
\newcommand{\mn}[1]{$\mathtt{#1}$}			
\newcommand{\um}{\ifmmode \umu \text{m} \else $\umu$m\xspace \fi}		

\title[Global dust attenuation in disc galaxies]{Global dust attenuation in disc galaxies: strong variation with specific star formation and stellar mass, and the importance of sample selection}

\author[B. M. Devour \& E. F. Bell]{
Brian M. Devour,$^{1}$\thanks{E-mail: bdevour@umich.edu}
Eric F. Bell,$^{1}$\thanks{E-mail: ericbell@umich.edu}
\\
$^{1}$Department of Astronomy, University of Michigan, 1085 South University Ave., Ann Arbor, MI 48109, USA\\
}

\date{Accepted 2016 March 30. Received 2016 March 28; in original form 2015 November 6}

\pubyear{2016}

\begin{document}
\label{firstpage}
\pagerange{\pageref{firstpage}--\pageref{lastpage}}
\maketitle

\begin{abstract}
We study the relative dust attenuation--inclination relation in 78,721 nearby galaxies using the axis ratio dependence of optical--NIR colour, as measured by the Sloan Digital Sky Survey (SDSS), the Two Micron All Sky Survey (2MASS), and the Wide-field Infrared Survey Explorer (\emph{WISE}). In order to avoid to the greatest extent possible attenuation-driven biases, we carefully select galaxies using dust attenuation-independent near- and mid-IR luminosities and colours. Relative $u$-band attenuation between face-on and edge-on disc galaxies along the star forming main sequence varies from \tld0.55 mag up to \tld1.55 mag. The strength of the relative attenuation varies strongly with both specific star formation rate and galaxy luminosity (or stellar mass). The dependence of relative attenuation on luminosity is not monotonic, but rather peaks at $M_{3.4\um} \approx -21.5$, corresponding to $M_* \approx 3\times 10^{10}M_{\sun}$. This behavior stands seemingly in contrast to some older studies; we show that older works failed to reliably probe to higher luminosities, and were insensitive to the decrease in attenuation with increasing luminosity for the brightest star-forming discs. Back-of-the-envelope scaling relations predict the strong variation of dust optical depth with specific star formation rate and stellar mass. More in-depth comparisons using the scaling relations to model the relative attenuation require the inclusion of star-dust geometry to reproduce the details of these variations (especially at high luminosities), highlighting the importance of these geometrical effects. 
\end{abstract}




\bibliographystyle{mnras}

\section{Introduction} \label{sec:intro}

The presence of dust strongly affects almost every observable property of gas-rich, star-forming galaxies in the optical, ultraviolet, and some infrared bands. Short-wavelength light from star forming regions is absorbed, heating the dust and effectively reprocessing this light into the infrared (see, e.g., \citealt{witt_gordon_00}; \citealt{charlot_fall_00}; \citealt{calzetti_01} for an overview). Dust not only absorbs but also scatters optical and ultraviolet light, reducing the emitted flux in some directions, while potentially even boosting it in others (e.g. \citealt{dejong_96}). The exact interplay of absorption and scattering depends sensitively on the relative geometry of stars and dust. This geometry is generally quite complex -- young stars and star-forming regions are often embedded in dense dust, different stellar populations often have different scale lengths and heights, and the dust distribution can also be highly structured and clumpy (see, e.g., \citealt{witt_gordon_00}; \citealt{holwerda_etal_07}; \citealt{schechtman-rook_etal_12}; \citealt{liu_etal_13}). As such, the total attenuation due to dust along any given line of sight -- defined as the sum total of the removal of light due to absorption and the scattering of light both into and out of the line of sight -- is difficult to {\it a priori} predict (see, e.g., \citealt{bruzual_etal_88}; \citealt{witt_gordon_00}; \citealt{jonsson_etal_06}; \citealt{steinacker_etal_13}).

The magnitude of dust attenuation and its effects on our inferences about galactic properties can be large. Averaged over the present-day galaxy population, roughly one third of all emitted starlight is absorbed by dust and reprocessed into the infrared \citep{soifer_neugebauer_91,driver_etal_12}. Almost all optically measured quantities of dust-rich disc galaxies are strongly affected by this dust attenuation -- examples of affected measurements include optical luminosities and colours (e.g. \citealt{tully_etal_98}; \citealt{maller_etal_09}; \citealt{masters_etal_10}), surface brightness profiles (e.g. \citealt{byun_freeman_kylafis_94}; \citealt{dejong_96}), and structural measurements such as half-light radii, B/T ratios, or \ser indexes \citep{pastrav_etal_13}. These dust effects influence scaling relations such as the Tully-Fisher relation (e.g. \citealt{tully_etal_98}), colour-magnitude relations  \citep{cho_park_09}, and stellar mass estimates (e.g. \citealt{bell_dejong_01}; \citealt{maller_etal_09}).

One possible approach to understand the impacts of dust on these measurements and scaling relations would be to model the effects of dust on the light from a galaxy using analytical prescriptions or numerical simulations (e.g. \citealt{bruzual_etal_88}; \citealt{disney_davies_phillipps_89}; \citealt{dejong_96}; \citealt{popescu_etal_00}; \citealt{tuffs_etal_04}; \citealt{jonsson_etal_06}; \citealt{steinacker_etal_13}). However, the attenuation from dust depends sensitively on the absorption and scattering properties of dust grains, the distribution and substructure of dust within a galaxy, and the distribution of stars, including the varying distributions of different populations of stars (e.g. younger and older) with different spectra (e.g. \citealt{witt_gordon_00}). This complexity means that modeling the dust attenuation, even in a simplified form, is a non-trivial task. Moreover, even for relatively sophisticated simulations, it is difficult to know for certain whether the dust properties, dust and star distributions, and other elements of the models actually match the physical properties of galaxies in the real universe. Overall, the distributions of stars and dust are not well characterized enough to yield robust predictions, so the predictions of these models remain useful, but primarily in a qualitative sense.

Accordingly, it is important to directly measure the effects of dust on the light distributions of galaxies. Historically, many methods have been used to explore the question of dust attenuation in disc galaxies. One method relies on observations of pairs of large disc galaxies that appear to overlap on the sky (e.g. \citealt{berlind_etal_97}; \citealt{keel_etal_14}). By measuring the change in the appearance of the two galaxies in the regions where they overlap and comparing it to their non-overlapping regions, using the assumption of symmetry one can measure the dust distribution in the foreground galaxy. Another method uses counts of distant background galaxies observed through the discs of large, nearby, face-on disc galaxies (e.g. \citealt{holwerda_etal_05}); such methods can be applied at very large distances from galaxies to reveal low-level extremely extended dust envelopes (e.g. \citealt{menard_etal_10}). Both of these methods are capable of giving fairly detailed insight into the dust distributions of their target galaxies -- in particular, they have established radial gradients in dust optical depth and that spiral arms are considerably richer in dust than inter-arm regions (e.g. \citealt{white_keel_conselice_00}; \citealt{holwerda_etal_05}) -- but are limited in terms of sample size.

The most statistically powerful method is to measure the dust-induced inclination dependence of parameters such as galaxy luminosities, colours, and structures (e.g. \citealt{valentijn_90}; \citealt{tully_etal_98}; \citealt{maller_etal_09}; \citealt{masters_etal_10}). Because it does not rely on chance alignments, large sample sizes spanning a range of galaxy properties are possible with modern sky surveys. As a consequence, it is well suited for measuring how dust attenuation varies among the galaxy population as a function of observable galaxy properties. This method addresses only the observed relative attenuation difference between edge-on and face-on galaxies. On one hand, this means that this methodology cannot correct galaxy properties to their `intrinsic' dust-free values without invoking models. On the other hand, this means that these measurements are robust and model-independent; they can be used to provide a useful check of more model-dependent dust correction methods that require assumptions about the stellar populations and attenuation curves of target galaxies. 

This technique has been employed many times in the past, with varying results. Optically-selected samples have proven challenging to analyze; the quantities used to select samples suffer from dust attenuation, and early results were contradictory (\citealt{valentijn_90}; \citealt{huizinga_vanalbada_92}; \citealt{giovanelli_etal_94}). Subsequent samples selected in the (nearly attenuation-independent) near-infrared (e.g. \citealt{tully_etal_98}) concluded that dust attenuation varies strongly with luminosity and is almost negligible for faint galaxies; this broad trend is now well-established (e.g. \citealt{maller_etal_09}; although \citealt{masters_etal_10} argue that attenuation decreases again somewhat towards the brightest luminosities). Dust attenuation appears to vary with other galaxy parameters, such as surface brightness (e.g. \citealt{dejong_lacey_00}) or morphology (e.g. \citealt{masters_etal_10}). Most such studies have chosen different samples, analysis methods, or parameters of interest, and these differences have made it difficult to develop a comprehensive overall picture of how dust content and dust attenuation (which are \emph{not} necessarily the same thing, as we explore later) vary among the diverse galaxy population. 

For example, (as explored later in this paper) \citet{tully_etal_98} and \citet{maller_etal_09} both measure attenuation in terms of change in optical-K band colour with axis ratio for galaxies of similar K band luminosity. Despite this similarity, and despite the fact that both parametrize variations in attenuation in terms of absolute magnitude, they recover differing results, both in predicting significantly different levels of attenuation for galaxies of similar luminosity, and more broadly in disagreeing on the strength of the luminosity dependence of attenuation by factors of 2-3. It is difficult to pin down the physical significance of these disagreements, however, due to the  differences between the studies -- are they related to \citeauthor{tully_etal_98}'s parametrization in terms of BRI magnitudes versus \citeauthor{maller_etal_09}'s use of K? Is part of the variation in attenuation being expressed in \citeauthor{maller_etal_09}'s additional parameter of \ser index?

In general, this lack of uniformity between different works makes it difficult to discern which parameters are most important in controlling dust attenuation. The broad outlines are visible, but the quantitative details are not. Many studies have focused on galaxy luminosity as the primary source of variation in attenuation. Yet other parameters such as star formation rate (e.g. \citealt{wild_etal_11}) or surface density (e.g. \citealt{dejong_lacey_00}) can be argued to be equally (if not more) fundamental, but comparison between current studies does not allow for a conclusive resolution of such questions. What is needed is an effort that addresses this by measuring the variation in dust attenuation as a function of many parameters simultaneously, using the same methodology and uniformly selected samples. This work represents the first part of such a study, exploring the variation in optical attenuation in the SDSS bands $ugriz$ as a function of \emph{WISE} parameters (scaling with stellar mass and star formation rate). In the future we will expand this by adding measurements of galaxy size, surface density, and structure to create a more comprehensive catalog of variation in attenuation with galaxy properties.

In addition to measuring the dependence of attenuation on stellar mass and star formation rate, our work focuses on the question of selecting samples in a uniform, unbiased manner. Since we cannot actually observe the same galaxy from multiple angles, measuring the inclination dependence of dust attenuation requires creating ensembles of galaxies which we believe to be essentially identical, save for their viewing angles. With such ensembles in hand, one can then study how the observable properties of these samples vary as a function of inclination to robustly measure the relative attenuation from edge-on to face-on orientation. Properly assembling these samples is the key challenge of this method, as any inclination-dependent variation in inherent properties among the members of the sample will bias the resulting attenuation measurements.

This is now possible thanks to advances in wide-area near- and mid-IR surveys, for example, the Two Micron All-Sky Survey (2MASS; \citealt{skrutskie_etal_06}), the near-IR Large Area Survey from the UKIRT Infrared Deep Sky Survey (UKIDSS-LAS; \citealt{lawrence_etal_07}), and especially the all-sky mid-IR coverage from 3.4\um to 22\um provided by the Wide-field Infrared Survey Explorer (\emph{WISE}; \citealt{wright_etal_10}). Near-IR wavebands such as K or 3.4\um provide measurements of total stellar luminosity (and stellar mass to within factors of two; \citealt{bell_dejong_01}; \citealt{cluver_etal_14}; \citealt{meidt_etal_14}) and are well-resolved enough to measure galaxy structures. Mid-IR bands such as 12\um or 22\um measure hot and warm dust emission that correlates with star formation activity. Accordingly, these surveys allow us to probe important physical parameters of galaxies in a manner which is not affected by dust attenuation.

It is an important point that the analysis technique presented here does not, and is not intended to, correct galaxy properties to their intrinsic, dust-unaffected states. Rather, this analysis corrects for the \emph{additional} effects of inclination -- in effect, it attempts to correct all galaxies to their \emph{face-on} states, rather than their \emph{dust-free} states. This is desirable for a number of reasons. First, for the purposes of classifying galaxies into different categories or comparing them to each other, this correction is usually sufficient -- for these purposes, the intrinsic dust-free properties are not relevant, and the simpler analysis offers less scope for the introduction of systematic error. More importantly, though, this restriction means that the technique requires no assumptions about the intrinsic physical properties of the galaxy such as dust attenuation curves, star/dust geometry, stellar populations, etc. Correcting face-on to dust-free invariably introduces model-dependent variation, and it is not immediately self-evident which models (if any) represent the correct approach for which galaxies. 

The plan of this paper is as follows. In \S \ref{sec:data} we discuss and carefully select the most appropriate catalog SDSS and WISE flux measurements, and critically examine available axis ratio measurements. We explain how we frame our sample selection in terms of \emph{WISE} 3.4{\um} luminosities ($M_{3.4\um}$) and 12{\um}--3.4{\um} colour ([12]--[3.4]), construct a sample which is independent of inclination, and discuss the demographics of our sample in the \emph{WISE} color--luminosity plane in \S \ref{sec:samp_demo}. Our analysis method and measurements of relative attenuation as a function of inclination are presented in \S \ref{sec:results}, and we compare these results in detail with two important previous works in \S \ref{sec:tully_maller}. In \S \ref{sec:discussion} we compare our results more briefly with some other notable previous works and explore the implications of a simple scaling relation model of our results. We conclude in \S \ref{sec:conclusions}. As one of the goals of this paper is to illustrate carefully the influence and importance of input data choices and sample selection on studies such as these, parts of this paper may be less relevant to some readers. As such, readers less interested in technical and historical details are invited to focus on Sections \ref{sec:samp_demo}, \ref{sec:results}, \ref{subsec:scaling} and \ref{sec:conclusions}. All magnitudes throughout are presented in the AB system \citep{oke_gunn_83}, all logarithms are base-10, and where necessary we assume a cosmology of $\Omega_{M} = 0.3$, $\Omega_{\Lambda} = 0.7$, and $H_{0} = 70 \text{ km s}^{-1} \text{ Mpc}^{-1}$.

\section{Data} \label{sec:data}

Our goal is to measure the difference between edge-on and face-on attenuation in the optical SDSS $ugriz$ passbands as a function of parameters drawn from longer-wavelength datasets (\emph{WISE} and 2MASS) that scale with stellar mass and dust-enshrouded star formation rate. Accordingly, we choose for analysis a sample of galaxies drawn from the Sloan Digital Sky Survey (SDSS) DR10 (\citealt{eisenstein_etal_11}, \citealt{ahn_etal_14}), cross-matched with data from the Wide-field Infrared Survey Explorer (\emph{WISE}) All-Sky Data Release \citep{wright_etal_10}. We use axis ratio measurements derived from the SDSS bulge/disc decompositions of S11 as the best available measurements of galaxy inclination, and for some of our sample diagnostics we make use of data from the the 2-Micron All Sky Survey Extended Source Catalog (2MASS XSC; \citealt{skrutskie_etal_06}). After all cross-matching and cuts, we have a sample of approximately 80,000 galaxies of all types with reliable axis ratios, SDSS optical photometry, and \emph{WISE} dust- and inclination-independent NIR-MIR photometry.

As noted below, basic catalog measurements often suffer from issues which make them less ideal for our purposes. Nonetheless, we use these catalog measurements whenever possible rather than derived quantities for two main reasons -- first, to avoid any hidden dust-dependent systematic errors in more complex model-derived quantities, and second, to enable easy use of our work by others.

\subsection{Photometric Catalog Data}

Our initial sample selection is from the SDSS, drawing from the Galaxy view of the $\mathtt{PhotoPrimary}$ table and selecting objects with clean photometry as defined by the flag CLEAN = 1 and associated spectral data in the $\mathtt{SpecObj}$ table. The SDSS provides the optical magnitudes we use to probe dust attenuation, but the reason to start with this base sample is that the SDSS is the source of the spectroscopic redshifts we use to calculate absolute magnitudes. The only SDSS-based selection criteria applied to this inital sample is a limit of $m_{r} \leq 17.7$ mag, to ensure clean spectral data. (We are cognizant that optical selection limits may introduce biases, and as described in \S\ref{subsec:i_indep} we use an additional redshift selection in our analysis which controls for this effect.) There is also roughly 6 per cent incompleteness in the SDSS spectroscopic sample due to fiber collisions \citep{strauss_etal_02}, but as this selection is geometric rather than dependent on galaxy properties it will not introduce any biases in our sample. This returns a sample of 513,597 galaxies, which, after later cross-matching with \emph{WISE}, results in a final sample of 78,721 galaxies.

In this subsection we detail the various photometric measurements available in our catalogs, and explain our reasoning for choosing the specific quantities we use. The choice of photometric parameters has a notable quantitative effect on our analysis, though not enough to make a qualitative difference in our final conclusions (see discussion in \S\ref{subsec:errors}).

\subsubsection{WISE} \label{subsec:wise}

In our sample, we use \emph{WISE} infrared data to provide inclination-independent measures of galaxy properties. We take our initial catalog of SDSS galaxies and cross-match it with \emph{WISE}, treating any \emph{WISE} source falling within 5 arcsec of the centre of an SDSS galaxy as a match. The SDSS-based limit of $m_{r} \leq 17.7$ noted above is significantly more restrictive than the \emph{WISE} detection limits, and so this cross-match is able to match the vast majority of our sources, with the initial cross-matching resulting in a sample of 509,305 galaxies with SDSS and \emph{WISE} catalog properties. However, as explained below, the number of galaxies with photometry measured in the most useful manner is smaller, at 78,721.

From \emph{WISE}, we extract galaxy apparent magnitudes $m_{\text{w}1}$, $m_{\text{w}2}$, $m_{\text{w}3}$, $m_{\text{w}4}$ in the \emph{WISE} W1 (3.4\um), W2 (4.5\um), W3 (12\um), and W4 (22\um) bands. The foreground extinction in these bands is essentially zero, so no extinction corrections are necessary, but K-corrections are applied using the IDL \mn{kcorrect} package \citep{blanton_roweis_07}. Absolute magnitudes $M_{3.4\um}$, $M_{4.5\um}$, $M_{12\um}$, $M_{22\um}$ are calculated using spectroscopic redshifts from the SDSS $\mathtt{SpecObj}$ table, and the results are converted to the AB photometric system.

However, there are a variety of different magnitude measurements given in the \emph{WISE} catalog, and not all of them are suitable for our purposes. The primary issue is that most \emph{WISE} photometry is optimized for measuring the brightness of point sources.

\begin{figure*}
\begin{center}
\includegraphics{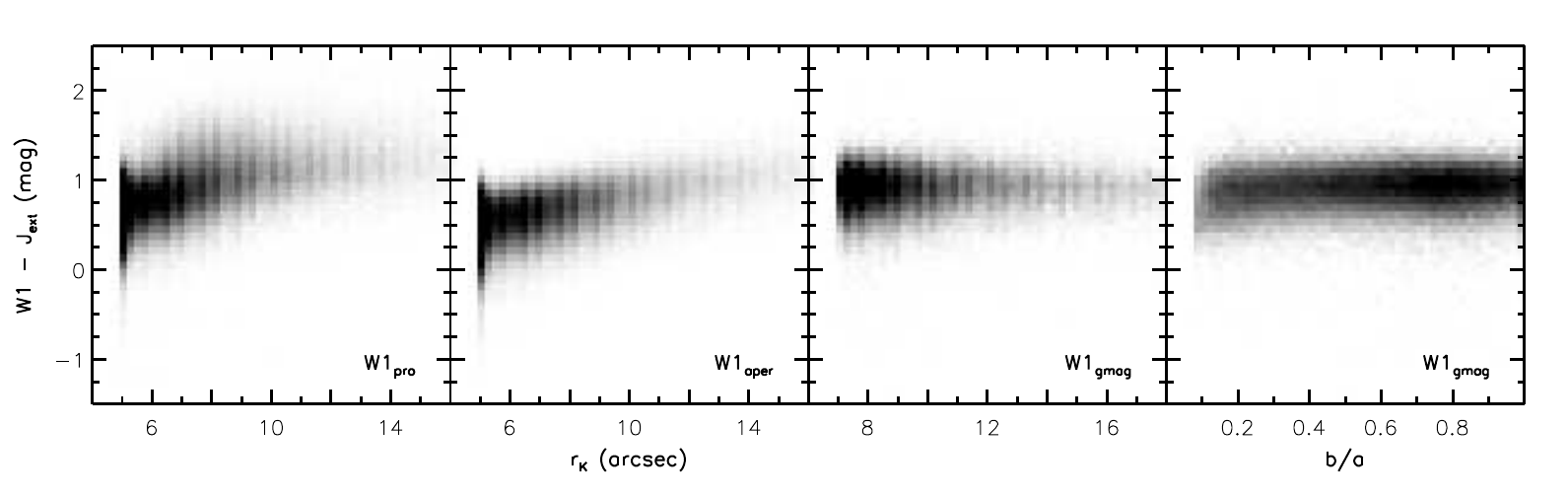}
\caption{Left to right: Panels 1-3: \emph{WISE} W1 profile-fit magnitude, \emph{WISE} W1 standard aperture magnitude, and W1 gmag, all minus 2MASS extended $J$ magnitude, as a function of 2MASS $K$-band radius. Panel 4: \emph{WISE} W1 gmag minus 2MASS extended $J$ magnitude, as a function of axis ratio.
\label{fig:wise_ap}}
\end{center}
\end{figure*}

Profile fit photometry fits the \emph{WISE} PSF to the observed profile of a source, assuming that the source is unresolved (\mn{wXmpro} in \emph{WISE}, where X is a numeral 1--4 referring to the four \emph{WISE} bands). However, most of the galaxies in our sample have sizes which are similar to or larger than the 6 arcsec FWHM of the W1--3 band \emph{WISE} PSF. Since the profile fit photometry assumes an unresolved source, this results in substantial systematic errors in magnitudes and colours as a function of apparent galaxy size. This is shown in the first panel of Fig.\ \ref{fig:wise_ap}, where the W1 profile fit magnitudes are compared to the 2MASS extended source $J$ magnitudes (a measurement much less susceptible to this form of error) as a function of 2MASS $K$ band radius.\footnote{2MASS radii and extended source magnitudes are measured using properly-scaled elliptical apertures, mitigating these sorts of errors. $J$-band photometry is chosen because of its high S/N assisting in capturing the outskirts of a galaxy, while $K$-band radii are measured using the longest wavelength available.} It can be seen that the \emph{WISE} magnitudes are a strong function of radius. Additionally, the \emph{WISE} PSF varies in size in different bands, and the apparent size of the galaxy itself may also vary due to population gradients, giving further scope for systematic error in various difficult-to-predict forms.  Therefore, the profile fit photometry is not suitable for our use.

Standard aperture photometry returns the flux within a `standard' aperture (8.25 arcsec radius for W1--3, 16.5 arcsec radius for W4), corrected for the curve-of-growth of the PSF again assuming an unresolved source (\mn{wXmag} in \emph{WISE}). This type of photometry suffers from many of the same problems as the profile fit photometry. Since the aperture correction assumes an unresolved source while our galaxies are generally resolved, the fluxes will be underestimated, particularly for larger galaxies. Again, this is shown by comparison to 2MASS extended source $J$ magnitudes in the second panel of Fig.\ \ref{fig:wise_ap}. The standard aperture colours are also affected by colour gradients due to the aperture preferentially sampling the core rather than the outskirts. The aperture photometry is also vulnerable to contamination to some degree, though to a relatively limited extent due to the fairly small aperture. For these reasons, the standard aperture photometry is also not suitable for our work.

Finally, multi-aperture photometry returns the flux within a series of apertures with radii ranging from 5.5--24.75 arcsec (W1--3) or 11.0-44.0 arcsec (W4), with no further correction (\mn{wXmag\_Y} in \emph{WISE}, where Y is a numeral 1--8 referring to the 8 different aperture sizes). The multi-aperture photometry is theoretically not subject to the radius-dependent errors of the previous two varieties, because an appropriately sized aperture can be chosen to correctly capture our extended sources. However, the multi-aperture photometry is significantly more vulnerable to contamination as the size of the aperture increases. Additionally, the multi-aperture photometry uses circular apertures, which further broadens the scope for systematic errors with galaxy size and axis ratio. For these reasons, the multi-aperture photometry is also less than ideal for our purposes. 

The most useful \emph{WISE} photometry for galaxy magnitudes is the 2MASS-association-based elliptical aperture photometry (\mn{wXgmag} in \emph{WISE} -- referred to henceforth as gmag). \emph{WISE} sources are internally cross-matched with the 2MASS XSC, and sources which lie within 2 arcsec of a XSC source are assumed to be that same object. In addition to the standard photometry, these sources are measured using an elliptical aperture whose radius, axis ratio, and position angle are derived from the 2MASS $K$ band `standard' elliptical aperture for that source. This provides a non-point source magnitude measured in an aperture whose size and shape are scaled appropriately based on non-optical measurements of that source, thus avoiding most of the systematic errors present in the profile-fit, standard aperture, and multi-aperture photometry. This can be seen in the remaining two panels of Fig.\ \ref{fig:wise_ap}. In the fourth panel we compare the gmags to the 2MASS extended source $J$ magnitudes as a function of radius in the same manner as before, and we find that the gmags do not have any significant dependence on radius, unlike the profile fit and standard aperture photometry. Similarly, in the fourth panel, we see via the same comparison that the gmags also do not show any significant inclination dependence. This is the photometry we use for our work.

The gmags are the best overall magnitude available in the \emph{WISE} catalog, but using these does come with some significant drawbacks, the most obvious of which is sample size. While we have 509,305 galaxies with basic SDSS+\emph{WISE} photometry, only 78,721 of them have gmags measured in both W1 and W3 due to the requirement of cross-matching with 2MASS. The lower sensitivity of 2MASS eliminates some of the faintest galaxies, and more are eliminated due to this photometry's requirement of $r \geq 7$ arcsec in 2MASS $K$-band. Also, the 2MASS $K$-band `standard' aperture that these magnitudes are based on is defined by the 20 mag arcsec$^{-2}$ surface brightness isophote, which does not capture the faint outskirts of a galaxy and thus underestimates the total flux. Therefore, while the colours measured using these magnitudes are quite accurate, the total luminosities may be an underestimate. However, it is important to note that the purpose of our selection criteria is to divide our sample into bins of essentially identical galaxies. For this purpose it does not actually matter whether the luminosities are accurate -- merely that, if they are biased, that this bias affects all galaxies equally and does not depend on the parameters of the galaxies involved. This appears to be the case for these magnitudes, as we find no systematic trend in the gmag measurements with properties such as radius. Therefore, while this choice of photometry does limit our sample size and comes with some caveats, the resulting sample is the only one that is free from all of the various biases described above.\footnote{There is one further set of magnitudes we might choose to use -- the SDSS/\emph{WISE} forced photometry of \citet{lang_etal_14}. In principle this catalog could combine good extended source photometry with larger sample sizes. In practice, however, in order to avoid selection biases we subject our sample to relatively restrictive volume limits as detailed in \S\ref{subsec:i_indep}, so the sample size advantage of using \citeauthor{lang_etal_14}'s magnitudes is limited. Additionally, since this is a set of model magnitudes based on optically-derived SDSS parameters, there are several potential systematics that we do not yet feel we understand well enough to be comfortable using these magnitudes. This decision may be revisited in our future work.}

\subsubsection{SDSS} \label{subsec:sdss}

The SDSS optical catalog data provide the dust-sensitive measurements we use to quantify the dust attenuation in our sample. From the SDSS, we extract the apparent magnitudes $m_{u}$, $m_{g}$, $m_{r}$, $m_{i}$, $m_{z}$ in the 5 SDSS bands. These magnitudes are corrected for foreground Galactic extinction using the extinction values given in the \mn{Extinction\_X} (where $\text{X} \in \{ \text{u,g,r,i,z} \}$ refers to one of the five SDSS bands) columns of the SDSS and are K-corrected using the IDL \mn{kcorrect} package \citep{blanton_roweis_07}. As with \emph{WISE}, galaxy redshifts $z$ are adopted from the \mn{SpecObj} table and are used to calculate the absolute magnitudes $M_{u}$, $M_{g}$, $M_{r}$, $M_{i}$, $M_{z}$.

As with \emph{WISE}, there are several different magnitudes given in the SDSS catalog, and like \emph{WISE}, many of the most commonly used are not suitable for our purposes.

Petrosian magnitudes (\mn{petroMag\_X} in SDSS) are measured in an aperture whose size is defined as the radius at which the local surface brightness in the $r$ band drops to a given fraction of the average surface brightness interior to that radius. While this includes the majority of a galaxy's light, it does not capture the full extended flux of a galaxy, and the extent to which the flux is underestimated depends on light profile shape (larger for light profile shapes with larger wings, such as the \dev profile). Also, since the Petrosian radius is defined in terms of surface brightness in circular annuli, this raises the possibility of axis ratio-dependent effects. Given that we have a wide range of galaxy morphologies in our sample and that our analysis relies on inclination dependence of galaxy magnitudes, we do not use Petrosian magnitudes.

In addition to the Petrosian magnitude, there are two types of model magnitude. The standard model magnitudes (\mn{modelMag\_X} in SDSS) are based on fitting the light profile of each galaxy in the $r$ band with both an exponential and a \dev profile, choosing the better profile fit, and extrapolating that profile fit in each band (appropriately scaled) to estimate the galaxy's total flux. This model magnitude is good for tasks that require accurate measurement of SDSS colours, since the use of the $r$-band fit means that flux is measured in an equivalent way in all bands. However, this magnitude has the limitation that it is derived solely from the better fitting of the exponential \emph{or} the \dev fits, even if the galaxy is actually best fit by a combination. This leads to inconsistent results when applied to galaxies which have intermediate light profiles (bulgy spirals or discy ellipticals). This can cause systematic errors in SDSS--NIR colours with morphology, as well as leading to discontinuities in magnitude measurements for the population of galaxies which are almost equally well fit by either profile. Since our work includes galaxies with a wide range of morphologies and we are most concerned with the stability of SDSS-NIR colours rather than intra-SDSS colours, we also do not use standard modelMags.

Rather, we use the composite model magnitudes (\mn{cModelMag\_X} in SDSS). These magnitudes are based on the same exponential and \dev fits to each galaxy's light profile, but gives a magnitude based on the linear combination of the two that best fits the galaxy's profile rather than simply using only the better of the two fits. This provides superior photometry for the large number of galaxies which are not well fit by a pure exponential or \dev profile. This fit is also done individually in each band rather than using the $r$-band fit for all bands, which sacrifices some accuracy for intra-SDSS colour measurements but further improves the quality of the photometry for comparisons to NIR data. Therefore, the composite model magnitudes are the best choice for our work.

\subsection{Measures of projected axis ratio} \label{subsec:ab}

\begin{figure}
\begin{center}
\includegraphics{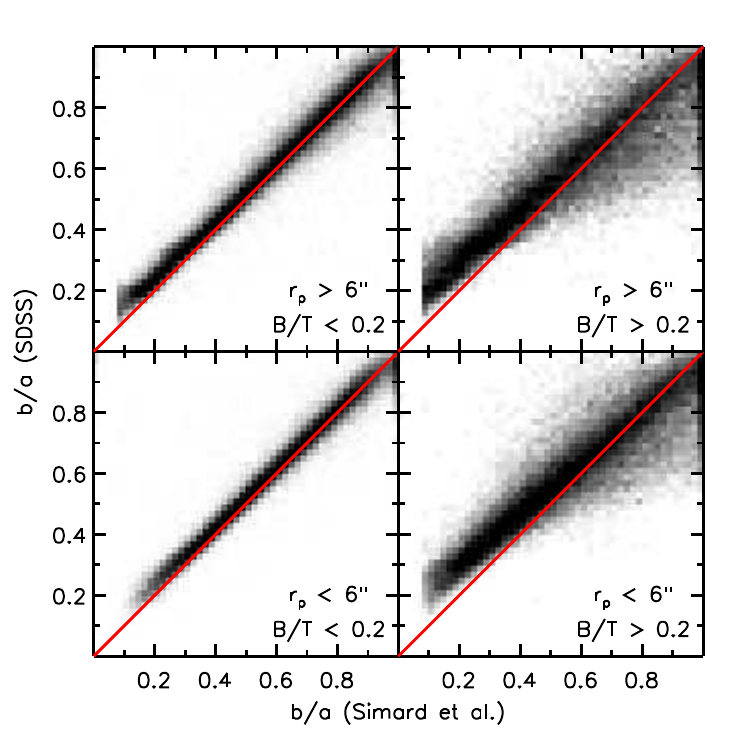}
\caption{Comparison of axis ratio measurements from S11 (horizontal axis) and SDSS (vertical axis) for disc galaxies, divided by B/T ratio and r-band angular size. In each panel the 2-d histogram shows the distribution of galaxies within that bin of B/T and angular size, and deviations from the diagonal 1-to-1 line reflect differences in the axis ratio measurements assigned by these two datasets.}
\end{center}
\end{figure}

We wish to measure the change in observed galaxy properties as a function of disc inclination $i$. However, inclination $i$ is not an immediately observable property -- rather, we observe galaxy axis ratios $a/b$, measured using isophotal fits or similar methods. The connection between galaxy inclination -- the fundamental physical property we wish to exploit -- and observed axis ratio depends on a host of other effects, including galaxy morphology differences such as intrinsic asymmetry and bulge size, photometric effects such as surface brightness limits and seeing, and even the very dust attenuation we wish to study. Therefore, the choice of inclination metric is an important one, and indeed has a significant quantitative and qualitative effect on the results (see discussion in \S\ref{subsec:errors}).

If our goal is to avoid visual wavelength measurements to reduce the risk of dust-induced biases, a natural choice of inclination metric would seem to be the 2MASS axis ratios. In the $K$ band, for example, these would be essentially unaffected by dust attenuation. However, the relatively low resolution of 2MASS imagery renders these measurements unsuitable for our purposes. In the online 2MASS documentation, \citet{cutri_etal_06} caution that the various galaxy shape measurements become unreliable for objects smaller than 10 arcsec, which encompasses roughly 70 per cent of our sample. Additionally, since these are isophotal axis ratios they will suffer from biases due to the presence of bulges, which we would prefer to avoid. And so, despite the desirability of a non-optical inclination metric, we do not use the 2MASS axis ratios.

The SDSS provides axis ratios based on their model magnitude exponential and \dev profile fits. These fits are convolved with the SDSS PSF to account for seeing, but as we show below, this correction is possibly incomplete. Additionally, since they are fit with either a pure exponential or a pure \dev profile, they cannot account for the presence of both a bulge and a disc in the target galaxies, which results in biased axis ratios. Finally, like the standard model magnitudes, the model axis ratios will also be subject to discontinuous behavior for the population of galaxies which are almost equally well fit by either profile. For these reasons, we do not use the SDSS axis ratios. 

Rather, for our analysis we adopt the $r$-band disc inclinations from the two-component \ser models \citep{sersic_63} of \citet{simard_etal_11} (hereafter S11). S11 perform two-dimensional bulge-disc fits to the $g$ and $r$-band images for galaxies in the SDSS, accounting for the effects of seeing. For our purposes, a key advantage is that for a disc galaxy such a measurement should account for the changes in $b/a$ expected from both the addition of the bulge and the convolution of a thin disc with a (much rounder) PSF; both effects would tend to drive a direct measurement of $b/a$ to larger values, impacting analyses such as ours. Thus, using this measurement should return more accurate disc axis ratios for disc galaxies with significant bulge components.

Of course, these fits and axis ratios do not operate in quite the same way for bulge-dominated galaxies without significant disc components. As discussed in S11, in such cases the disc component of the bulge-disc model instead fits to the outer, flatter portions of the bulge-dominated galaxy's light profile. In this case the axis ratio returned is obviously no longer a `disc' axis ratio; however, it is still a reliable measurement of the shape of the outer isophotes. This is fortuitous from our perspective, as it means that this single measurement returns accurate axis ratios for disc-dominated galaxies and reasonable axis ratios for bulge-dominated galaxies. While the brightest bulge-dominated galaxies are usually triaxial in shape, and thus not appropriate targets for our analysis method since their axis ratio distributions do not reflect variations in inclination, this does allow us to probe the inclination dependence for the intermediate populations of bulgy galaxies which tend to be shaped like oblate spheroids \citep{vanderwel_etal_09}.

Figure 2 shows a comparison between $b/a$ values derived from S11's disc inclinations and those derived from the SDSS models, illustrating the differences between these measurements.  This includes disc galaxies only (for our purposes, this is roughly defined as \emph{WISE} $M_{12\um} - M_{3.4\um}$ colour $< \tld0.5$ -- see discussion in \S\ref{subsec:demo}), and this sample is divided by angular galaxy size ($r$-band petrosian radius) as measured by SDSS and bulge/total ratios as measured by S11 (both of these parameterizations are potentially dust, and therefore inclination, dependent, but this division of the sample is purely for illustrative purposes).

In the left column of highly disc-dominated galaxies, as may be expected the correspondence is overall quite good. However, deviations are apparent for both the highest and lowest $b/a$ galaxies. For very high $b/a$ galaxies in the larger radius subsample, there is a slight trend for the SDSS $b/a$ measurements to fall below S11's. This is likely due to inherent asymmetries in the target galaxies, which can cause the measured $b/a$ value to fall short of unity even for a galaxy which is in truth being observed essentially face-on. S11's model fits appear to be less vulnerable to such effects than the SDSS fits, which we speculate is due to the ability of S11's two-component fits to account for the presence of a bar or other non-axisymmetric bulge. This effect is less prominent for the smaller galaxies, likely because in these cases seeing will tend to blur out weaker asymmetries. At the other end of the scale, there is a small but notable trend for the SDSS $b/a$ measurements to report larger axis ratios than S11 for the most edge-on galaxies. This is likely due to some combination of seeing and bulge effects. For edge-on galaxies whose size on the sky is not greatly larger than the average PSF, their apparent minor axis size will be inflated to a greater relative extent than their major axis size, which will lead to $b/a$ measurements which are biased high if the PSF correction is insufficient. Additionally, as a galaxy becomes more inclined, the relative contribution to the projected axis ratio from any bulge component becomes larger. The SDSS single-component fits cannot account for this, but S11's two-component fits can. The fact that this effect is present at all for this disc-dominated subsample suggests that it may be a seeing effect. However, it is also present for both the large and small galaxies, which suggests a bulge effect. Therefore, it is unclear which of these two effects contributes more to the observed offset in this case.

In the right column of galaxies with more significant bulge components both of these effects become much stronger. We observe a much stronger trend for the SDSS $b/a$ measurements to be lower than S11's at high $b/a$. The strengthening of this trend with increasing B/T ratio suggests that it is indeed being driven by bars or other non-axisymmetric central features which S11's two-component fits can account for but the SDSS's single-component fits cannot. This effect is still slightly more prominent for larger galaxies, but unlike for the disc-dominated subsample it is also significant for the smaller galaxies. Similarly, the trend for the SDSS $b/a$ measurements to be larger than S11's for the most edge-on galaxies is also larger than in the disc-dominated sample. The seeing effect discussed above may still be in play, but it is apparent that the significant increase in deviation in this subsample is being driven by the increasing effects of bulges on the SDSS-measured axis ratios of edge-on galaxies. It is also notable that for this larger B/T subsample this effect is significant even at intermediate $b/a$ values -- the axis ratios measured by the SDSS are somewhat inflated by the presence of the bulge even for galaxies which are not being viewed completely edge-on.

The sample considered here focuses only on disc galaxies, but for our purposes the axis ratios of the highest B/T, bulge-dominated galaxies are not especially important, since these spheroid galaxies are typically dust-poor and the fact that they have only weak or nonexistent discs makes them inappropriate targets for our analysis technique. Therefore, the effects noted above are all strong reasons to adopt S11's inclinations rather than the SDSS axis ratios. It is important to note, however, that this does not mean that this is the perfect solution. The use of a model photometric fit always has the potential to introduce model-dependent biases, and this potential increases for more complex models such as S11's. Additionally, as mentioned one would ideally prefer to have NIR-derived inclination measurements rather than optical to avoid any potential dust effects. However, since the inclination measurements of these fits rely most strongly on light from the outskirts of the galaxies (the least dust-affected areas), these effects should not be large enough to compromise our analysis, and regardless these possible systematics are less problematic than the known systematics of the SDSS (and 2MASS) axis ratios. And so while in some senses S11's inclinations are not completely ideal and it is important to remember that there may be issues introduced by the use of this data set, we believe that they represent the best available measurement of inclination for our purposes.

\section{Sample Selection Criteria \& Population Demographics} \label{sec:samp_demo}

We wish to study attenuation by dust by examining the inclination dependence of the optical luminosities of galaxies as a function of galaxy properties. As discussed previously, the core challenge of any such study is the ability to identify and select samples of galaxies based on intrinsic, dust-independent properties.

In choosing a set of parameters for characterizing the properties of galaxies for a study of dust attenuation, there are a few criteria which we wish to reflect on. Most importantly, the parameters must be as dust and inclination independent as possible -- in practice, this will drive selection passbands to as long wavelengths as circumstances allow. It is useful to adopt a directly-observed parameter (e.g., an observed colour instead of a specific star formation rate) to avoid any hidden dependence of model-derived quantities on dust attenuation. Furthermore, it is very desirable that the properties in question are either observationally or theoretically expected to correlate with dust attenuation, at least broadly. Finally, it is helpful to the community to use properties that are widely available.  

\subsection{WISE absolute magnitudes and colours as valuable galaxy classification parameters} \label{subsec:samp}

We maintain that \emph{WISE} near and mid-infrared catalog absolute magnitudes and colours satisfy the above criteria. \emph{WISE} probes long-wavelength near and mid-infrared light which suffers from more than an order of magnitude less dust {\it attenuation} than the optical passbands. Our analysis in \S \ref{subsec:wise} illustrated that \emph{WISE} `gmags' appear to be unaffected by galaxy size and inclination. They are observational quantities, and are readily available for large samples of galaxies. In particular, we choose to sort galaxies as a function of their \emph{WISE} absolute 3.4\um magnitude and $12\um - 3.4\um$ colours. 

\begin{figure}
\begin{center}
\includegraphics{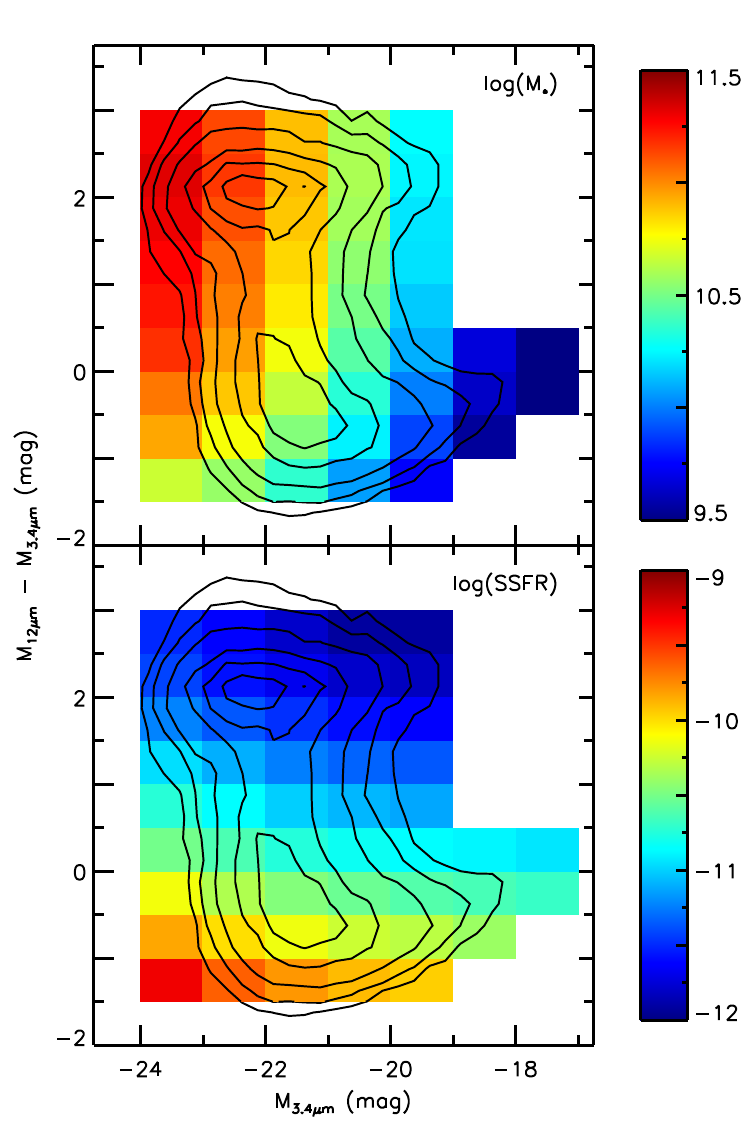}
\caption{Upper panel: stellar mass as a function of $M_{3.4\um}$ luminosity and [12]--[3.4] colour. The color scale shows $\text{log}(M_{*})$ in units of $M_{\sun}$. Lower panel: specific star formation rate as a function of $M_{3.4\um}$ luminosity and [12]--[3.4] colour. The color scale shows $\text{log}(\text{SFR}/M_*)$ in units of yr$^{-1}$. In both panels, background contours show the distribution of galaxies in this parameter space.  \label{fig:mstarssfr_ducks}}
\end{center}
\end{figure}

We choose to adopt \emph{WISE} 3.4\um absolute magnitude (henceforth referred to as $M_{3.4\um}$) as an observational proxy for stellar mass. Dust attenuation at 3.4\um is small, dust {\it emission} from PAHs and hot dust is moderate on galaxy-wide scales, and stellar M/L variations are expected to be modest $<0.3 \text{dex}$ (\citealt{meidt_etal_12,meidt_etal_14}, \citealt{cluver_etal_14}, see also Fig. \ref{fig:mstarssfr_ducks} and discussion below). Dust attenuation would {\it a priori} be expected to correlate with $M_{3.4\um}$ as both gas density and metallicity (which is expected to influence the dust-to-gas ratio) vary as a function of stellar mass (e.g. \citealt{tremonti_etal_04}). Indeed, previous works find a correlation between attenuation and $K$-band luminosity (e.g. \citealt{tully_etal_98}, \citealt{maller_etal_09}). 

In addition to $M_{3.4\um}$ luminosity, we choose also to select galaxies for study using their \emph{WISE} $12\um - 3.4\um$ colour (henceforth referred to as [12]--[3.4]). Again, the effects of dust attenuation on [12]--[3.4] colour are small, and this is a purely observational quantity, not dependent on dust attenuation sensitive models for sample classification. The [12]--[3.4] colour is expected to be a good proxy for dust enshrouded specific star formation rate (star formation rate per unit stellar mass; discussed further by \citealt{chang_etal_15}). Specific star formation rate is expected {\it a priori} to correlate with dust attenuation. Star formation rate should scale with gas density \citep{kennicutt_89,kennicutt_98} and therefore dust density; indeed, \citet{wild_etal_11} noted that attenuation curve shape varies strongly as a function of specific star formation rate. As noted above, $M_{3.4\um}$ luminosity correlates very strongly with stellar mass. Meanwhile, 12\um emission arises primarily from PAH molecules, small dust grains, and hot dust \citep{calzetti_13}. While a variety of heating mechanisms contribute to 12\um emission, the 12\um luminosity correlates to within a factor of two with the total infrared luminosity, which in turn gives insight into the dust enshrouded star formation rate \citep{papovich_bell_02,bell_03,wen_etal_14,chang_etal_15}. In practice, 12\um luminosity not only measures dust-enshrouded star formation, but for star-forming galaxies very closely tracks overall star formation rate as well. \citet{wen_etal_14} compare 12\um luminosity with Balmer-decrement extinction corrected H$\alpha$ luminosity (an alternate measure of star formation activity) for star-forming galaxies and find a tight relationship between IR and Balmer line brightness; estimating by eye from their fig. 3 suggests an RMS in the relationships of roughly 0.1 dex. Meanwhile, we adopt bins that are a half-magnitude (0.2 dex) wide in [12]--[3.4] -- thus, we are not attempting to divide our data more finely than the precision of this estimator. Therefore, we maintain that [12]--[3.4] is a robust and sufficiently effective diagnostic of star formation per unit stellar mass to meet our goals.

Figure \ref{fig:mstarssfr_ducks} shows estimated stellar mass and specific star formation rate for our sample, calculated from \emph{WISE} magnitudes and colors using the 3.4 and 4.6\um-stellar mass relation of \citet{cluver_etal_14} and the 12\um-star formation rate relation of \citet{wen_etal_14}. The contours show the overall galaxy distribution within the parameter space defined by $M_{3.4\um}$ luminosity and [12]--[3.4] colour, and the colour coding shows the average stellar mass or specific star formation rate at that location in parameter space. We can see that our interpretation of $M_{3.4\um}$ luminosity and [12]--[3.4] colour as proxies for stellar mass and specific star formation rate is, as a whole, reasonably justified. Stellar masses range from $\tld10^{9.5}$ $M_{\sun}$ at the very dimmest end of the star-forming main sequence to more than $10^{11}$ $M_{\sun}$ for the brightest quiescent galaxies. There is a tilt due to varying stellar mass-to-light ratios in the 3.4\um band between quiescent and star-forming galaxies (see \citealt{cluver_etal_14}), but overall, and in particular when considering galaxies at a fixed [12]--[3.4] color, $M_{3.4\um}$ serves as a reasonable proxy for stellar mass. Similarly, specific star formation rate varies from $\tld10^{-12}$ yr$^{-1}$ for the most quiescent galaxies to more than $10^{-9.5}$ yr$^{-1}$ for the most active galaxies. Again, while the same mass-to-light variations also produce a tilt in the distribution of specific star formation rate, both overall, and particularly when considering galaxies at a fixed $M_{3.4\um}$ luminosity, [12]--[3.4] color serves as a reasonable proxy for specific star formation rate.

These \emph{WISE} quantities satisfy our primary criteria in that they are long wavelength, dust attenuation insensitive, and widely-available observed quantities. Furthermore, as reasonable proxies for stellar mass and (dust-enshrouded) specific SFR, we expect them to capture some of the variation in dust attenuation from galaxy to galaxy. Nonetheless, there are doubtless other galactic parameters that should correlate with dust attenuation that one could select by. For example, the surface density of gas and stars should also correlate with attenuation (as argued e.g., by \citealt{dejong_lacey_00}), and bulge-to-total ratio (and more generally morphology) should affect attenuation by varying the star-dust geometry (e.g. \citealt{witt_gordon_00}). Unfortunately, such metrics are widely available only using dust attenuation- and inclination-dependent parameters (i.e., using half-light radii or morphologies from optical SDSS-derived catalogs; e.g. \citealt{blanton_etal_05}, S11) or are available for modest samples of very nearby galaxies (i.e., gas densities from HI or CO interferometry). In a subsequent paper, we will develop long wavelength-derived inclination-independent galaxy structure and morphology metrics and explore their correlation with dust attenuation; for our purposes in this paper, we focus on what is immediately available and study the variation of attenuation with \emph{WISE} catalog luminosities and colours.  

\subsection{Inclination Independence of Selection Limits} \label{subsec:i_indep}

\begin{figure}
\begin{center}
\includegraphics{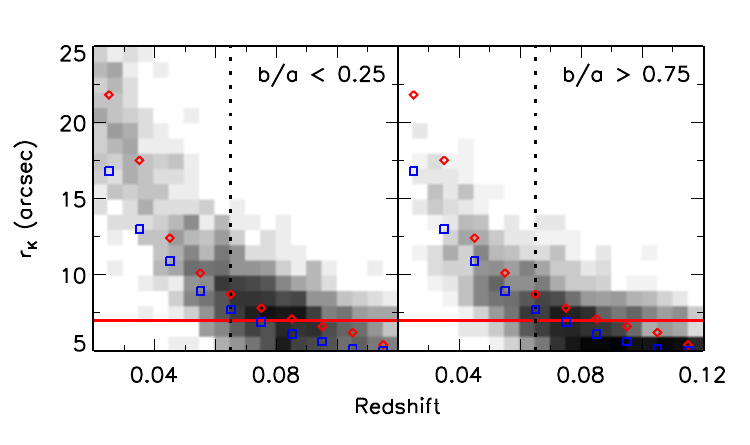}
\caption{Distribution of $K$-band radius as a function of redshift, for edge-on (left panel) and face-on (right panel) galaxies, in an example bin of parameter space ($-22<M_{3.4}<-21$, $-1.0<$ [12]--[3.4] $<-0.5$). Red diamonds and blue squares show the median of radius in redshift slices for the edge-on and face-on subsamples, respectively. The horizontal red line shows the 7 arcsec selection limit, and the vertical dashed line shows the adopted redshift limit in this bin.\label{fig:binradii}}
\end{center}
\end{figure}

While our individual subsample selection metrics are independent of inclination, it is important that our overall sample selection limits also have this property. Our initial sample is subject to two potentially problematic selection limits which we consider.

First is the limit of $m_r \leq 17.7$ mag in the SDSS, which is required to ensure clean spectroscopic data. Since this is an optical selection, for dusty disc galaxies this limit will introduce a bias against faint edge-on galaxies within each bin, as these galaxies may drop below the magnitude limit due to attenuation while their intrinsically equally-bright but less attenuated face-on counterparts do not.

Second, due to the required cross-match with 2MASS, the \emph{WISE} gmag photometry is subject to an apparent radius limit of $r \geq 7$ arcsec in the 2MASS $K$ band. This is an isophotal radius, so it is affected by the projected surface brightness of the target galaxy. Since galaxies are essentially transparent in $K$ band, their surface brightness will increase with inclination as the same amount of light is concentrated in a smaller area on the sky. Thus, for intrinsically identical galaxies, the measured isophotal radius will be increased for galaxies viewed closer to edge-on. This induces a bias which acts in the opposite sense as the SDSS apparent magnitude limit. In this case, edge-on galaxies which otherwise would lie just below the radius limit have their measured radius increased due to their inclination by enough to rise into the sample, while their intrinsically similar face-on counterparts remain below the cutoff.

Both of these biases are potentially problematic. Since our method relies on measuring the inclination dependence of dust-affected optical measurements in a subsample of intrinsically similar galaxies, it is required that the high-inclination and low-inclination components of that subsample are actually representative of the same underlying population. Thus, selection effects that act to include or exclude galaxies based on their inclinations are obviously not desirable, and the fact that we have two different selection limits whose biases act in opposite directions and whose relative sizes are difficult to predict only further increases the potential confusion.

To control for these effects, we introduce a redshift limit within each bin of $M_{3.4\um}$ luminosity and [12]--[3.4] colour in order to preempt these biased selection limits with an unbiased one. Since both apparent brightness and angular size decrease with distance, an appropriately chosen redshift limit ensures that the bin acts as a proper volume-limited subsample by removing (in an unbiased manner) galaxies which lie near to these problematic selection limits.

In practice, we find that the radius limit is generally more significant than the magnitude limit. (That is, as redshift increases, the distribution of isophotal $K$-band radii of face-on galaxies within a bin approaches 7 arcsec before the distribution of apparent $r$-band magnitudes of edge-on galaxies approaches 17.7 mag. This radius limit obviously does not correspond to an exact magnitude limit, but of the galaxies with $r \geq 7$ arcsec, 93 per cent have $m_r \leq 17$ and 79 per cent have $m_r \leq 16.5$.) Thus, within each bin we determine the appropriate limit by finding the redshift at which the fraction of face-on galaxies whose angular sizes fall below the 7 arcsec limit crosses a threshold of 33 per cent -- in other words, this is the redshift at which face-on galaxies begin to disproportionately drop out of the sample compared to edge-on galaxies due to the radius limit.

This procedure is illustrated in Figure \ref{fig:binradii}, which shows the distributions of radius with redshift for face-on and edge-on galaxies in a single example bin within our parameter space. It is easy to see that the face-on galaxies are systematically smaller than the edge-on galaxies at a given redshift, illustrating the bias in radius measurements. However, by considering only galaxies whose redshifts are less than the indicated limit we ensure that all galaxies are included equally in the sample regardless of their inclination. (Similarly, one could determine an analogous redshift limit based on finding where the distribution of apparent $r$-band magnitudes of edge-on galaxies approaches the 17.7 mag limit, but, as noted above, in practice this is not necessary as a separate step since the radius-based redshift limit is smaller than the magnitude-based limit.) Therefore, by taking this redshift cut within each bin, we ensure that the inclination-dependent radius and magnitude selection limits do not introduce biases into our sample.

\subsection{Sample Population Demographics} \label{subsec:demo}

\begin{figure}
\begin{center}
\includegraphics{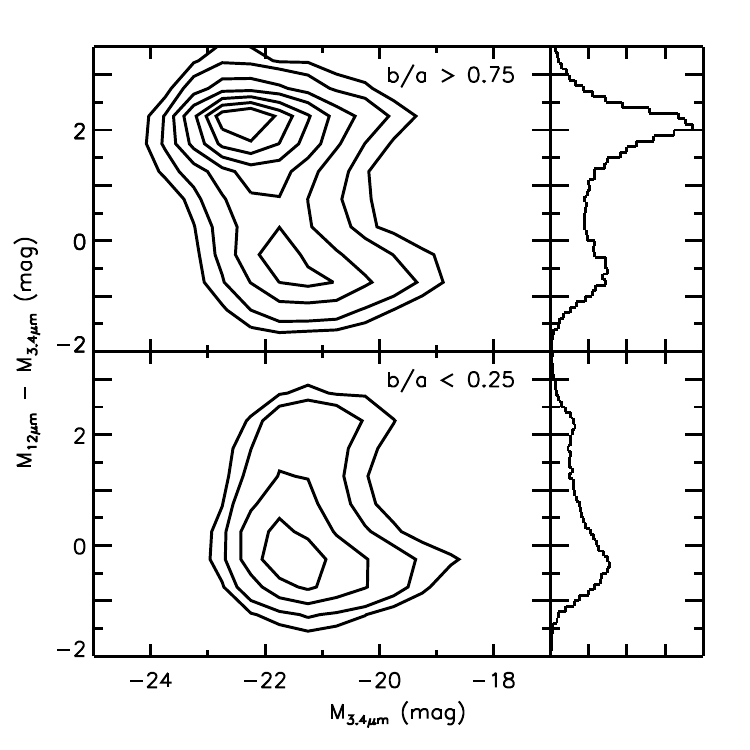}
\caption{Upper left panel: The distribution of galaxies with nearly circular shapes ($b/a>0.75$) in [12]--[3.4] colour as a function of $M_{3.4\um}$ absolute magnitude. Quiescent galaxies have [12]--[3.4] $> 0.5$ and relatively round shapes, and the star forming main sequence has [12]--[3.4] $\approx -0.5$. Upper right panel: The distribution of [12]--[3.4] colours for the $b/a>0.75$ subsample. Lower left panel: The distribution of galaxies with flattened shapes ($b/a<0.25$) in [12]--[3.4] colour as a function of $M_{3.4\um}$ absolute magnitude. Lower right panel: The distribution of [12]--[3.4] colours for the $b/a<0.25$ subsample.
\label{fig:contour_duck}}
\end{center}
\end{figure}

The joint selection on $M_{3.4\um}$ luminosity and [12]--[3.4] colour defines the parameter space for our analysis, and the properties of this parameter space also allow us to divide galaxies based on their morphology and isolate the disc galaxies we are interested in without relying on dust-affected morphology metrics such as SDSS $f_{DeV}$ or concentration. In Figure \ref{fig:contour_duck}, we show the distribution of our sample with respect to inclination in this parameter space.

The upper panel of Figure \ref{fig:contour_duck} shows the distribution of [12]--[3.4] as a function of $M_{3.4\um}$ absolute magnitude for a subsample with nearly round shapes (minor axis to major axis ratio $b/a>0.75$), corresponding to intrinsically round galaxies and/or low inclination discs. The lower panel shows the corresponding distribution for flattened galaxies with $b/a<0.25$, corresponding to highly inclined intrinsically flattened galaxies. The lower panel contains almost entirely disc galaxies; these have more negative [12]--[3.4] colours, having large amounts of 12\um emission relative to their 3.4\um emission. These galaxies have high (dust enshrouded) specific star formation rate. These star-forming galaxies cluster along a well-defined track in the $M_{3.4\um}$ vs.\ [12]--[3.4] parameter space -- the `star forming main sequence' -- due to the tight correlation between specific star formation rate and stellar mass (e.g. \citealt{salim_etal_07}; \citealt{noeske_etal_07}). Meanwhile, in the top panel we can see another prominent population of galaxies at less negative [12]--[3.4] colours. These galaxies have weak 12\um emission and therefore low star formation rates, and are mostly round, quiescent bulge-dominated galaxies. We can also see in the bottom panel a small number of low star formation rate elongated galaxies, which represent less common edge-on quiescent S0 galaxies.

The population of quiescent galaxies varies with axis ratio due to these galaxies' intrinsic round shapes, but these galaxies are not important for our purposes since they are expected to (and indeed do, see \S\ref{sec:results}) have little dust attenuation. Meanwhile, the population of star-forming disc galaxies is almost unchanged between the two bins of inclination, with a difference in median [12]--[3.4] colour of \tld0.1 mag. This is in contrast to optical studies such as \citet{tully_etal_98}, who found a difference in median $B-K$ colour of more than more than 1 mag between their highly inclined and low inclination samples. Therefore, these two parameters allow us to classify galaxies according to their stellar mass and specific star formation in an almost completely inclination independent manner.

\begin{figure*}
\begin{center}
\includegraphics{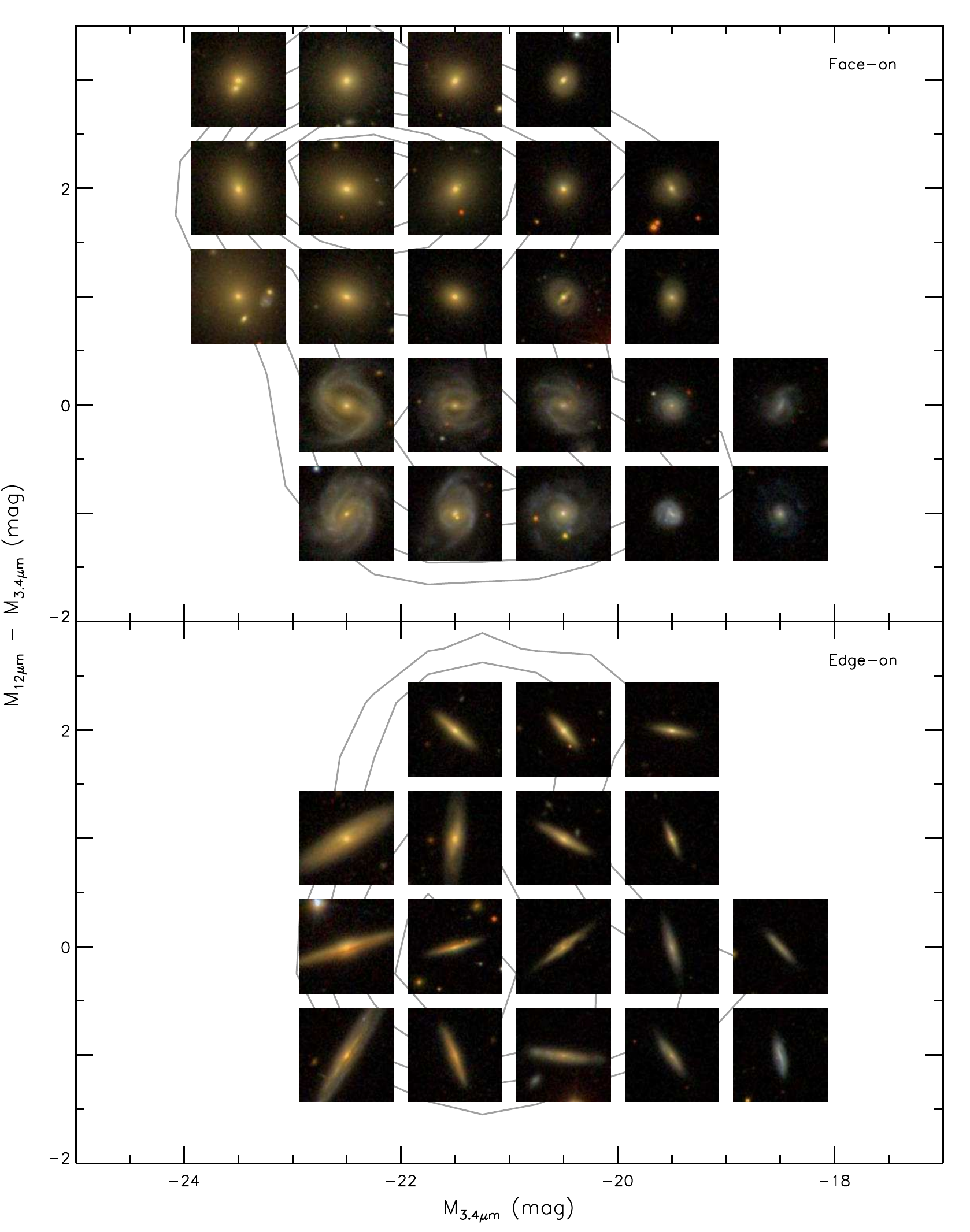}
\caption{Example SDSS postage stamp images for representative galaxies in our sample, placed according to their positions in our parameter space. Top panel shows face-on galaxies, bottom panel shows edge-on galaxies (the upper left-most bins are omitted in the bottom panel since there are very few highly-flattened galaxies in this region of parameter space). Background contours show overall galaxy distributions, as in Fig. \ref{fig:contour_duck}. All galaxies have $0.0333 < z < 0.05$ for size matching, except for the three bins with $-24 < M_{3.4\um} < -23$ in the upper panel, which extend to $z < 0.065$ instead due to fewer available nearby galaxies. \label{fig:btgs}}
\end{center}
\end{figure*}

These populations of quiescent, rounded galaxies and star-forming, flattened galaxies are illustrated visually in Figure \ref{fig:btgs}, which shows SDSS postage stamp images for representative galaxies throughout our parameter space. As in Figure \ref{fig:contour_duck}, the upper panel shows galaxies with large axis ratios, while the lower panel shows galaxies with small axis ratios. It is apparent that the region with [12]--[3.4] $> 0.5$ contains mostly bulge-dominated galaxies in the upper panel, though towards the bottom and right of this area some hints of discy features can be seen. In the lower panel, the upper left-most bins of parameter space are very sparsely populated and so not included, but the remainder of the region [12]-[3.4] $> 0.5$ appears to contain reddish, mostly featureless discs with unobscured bulges, suggesting S0 or similar galaxies. In both panels, the region [12]--[3.4] $< 0.5$ is obviously populated by discs. In the upper panel we can see that these discs contain obvious spiral structure, suggesting star-forming galaxies rather than quiescent S0s, and in the lower panel we can see that these discs appear significantly obscured when viewed edge-on (e.g. dust lanes or obscured bulges). Additionally, all of these galaxies were chosen from a restricted redshift range, so the size gradient from large galaxies on the left to small galaxies on the right is real and reflects the luminosity variation among our sample. All of these trends broadly confirm our physical interpretation of the divisions between types of galaxies in different regions of our parameter space.

\begin{figure*}
\begin{center}
\includegraphics{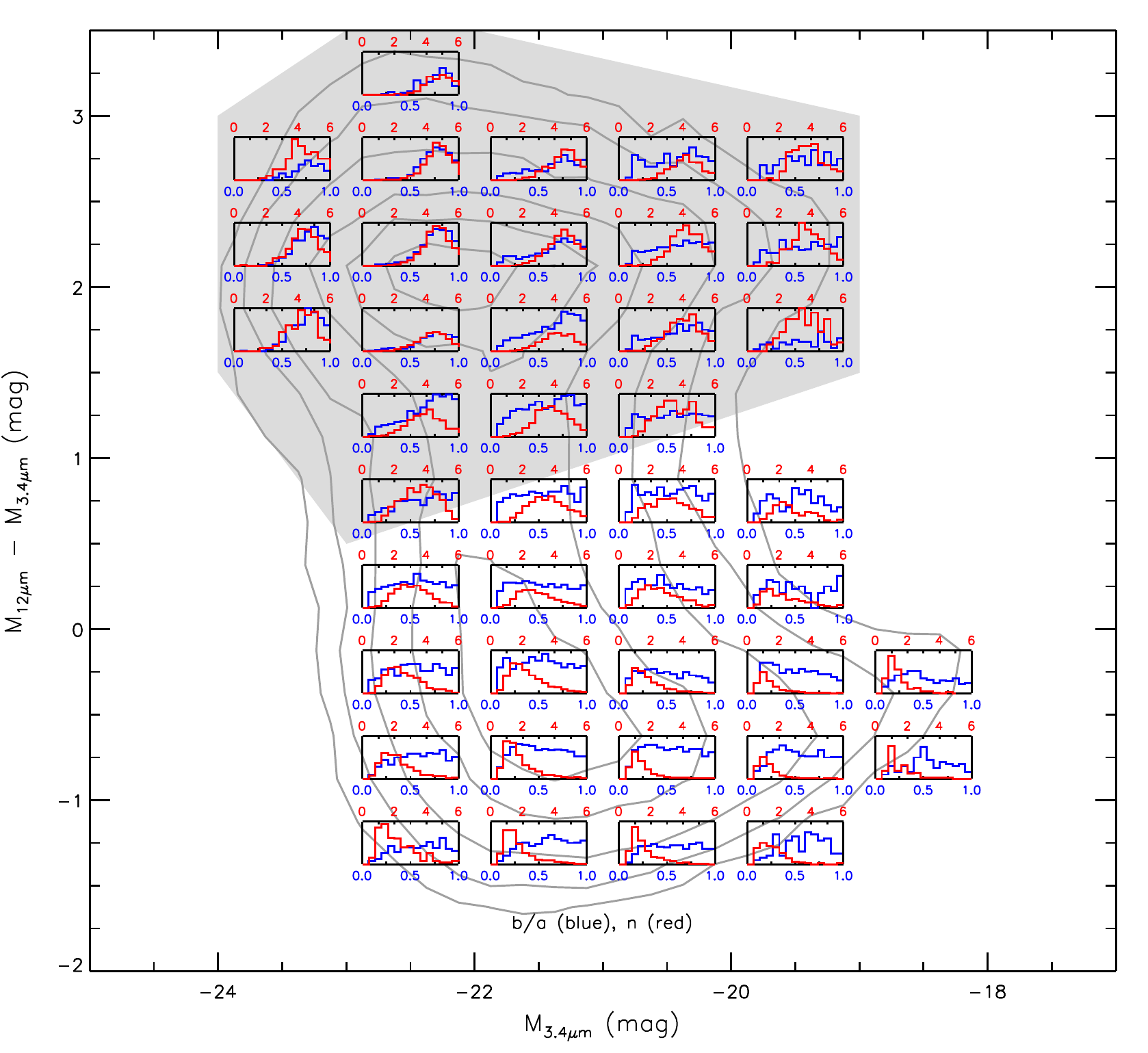}
\caption{Background contours show the distribution of galaxies in $M_{12\um} - M_{3.4\um}$ vs $M_{3.4\um}$ space. The inset panels show the distribution of axis ratios b/a (blue) and \ser index n (red) from S11 for galaxies at that location in $M_{12\um} - M_{3.4\um}$ vs $M_{3.4\um}$ space. Grey fill shows approximately the region of parameter space occupied by bulge-dominated galaxies. \label{fig:btba_duck}}
\end{center}
\end{figure*}

We further visualize the properties of this selected sample in Figure \ref{fig:btba_duck}, which shows the distributions of axis ratio and \ser index for our sample. We show the distribution of galaxies in the $M_{3.4\um}$ luminosity-[12]--[3.4] plane as background contours. Each overlaid panel shows the distribution of S11 axis ratio $b/a$ (in blue) and S11 \ser indexes $n$ (in red) at that location in parameter space (i.e. for subsamples selected to have narrow ranges in $M_{3.4\um}$ luminosity and [12]--[3.4]).

Towards the lower parts of the plot (towards negative [12]--[3.4] corresponding to star-forming galaxies), and on the right hand side even at [12]--[3.4] $\approx 0$, our sample has a nearly uniform distribution of $b/a$ -- a property of a sample of thin discs viewed from random directions. As outlined previously, this is an important property of our selection procedure: if photometry were inclination-dependent, it would drive galaxies towards preferential axis ratios in given colour bins. Such behavior is not seen for this combination of long-wavelength parameters and selection limits, illustrating that they are close to dust attenuation and inclination-insensitive and are well-suited for selection of samples with which we can study the effects of dust attenuation at {\it shorter} wavelengths.\footnote{This also can act as a diagnostic of axis ratio parameter -- the use of SDSS axis ratios does not give flat $b/a$ distributions regardless of selection.}

The upper left part of the distribution, with bright 3.4\um absolute magnitude and [12]--[3.4] $\approx 2$ corresponding to luminous quiescent galaxies, shows preferentially larger values of $b/a$. These galaxies are much rounder than the flattened star-forming population; detailed analysis indicates that these galaxies are likely to be triaxial spheroids (e.g. \citealt{tremblay_merritt_96}, \citealt{vincent_ryden_05}, \citealt{vanderwel_etal_09}). These galaxies violate our assumption that axis ratio $b/a$ reflects the inclination of a thin disc. We nonetheless analyze the run of colours with $b/a$ for this sample; if nothing else, it is important to verify that the change in colour with axis ratio is low for this sample of largely dust-free galaxies.

The red histograms show the distribution of \ser indices, measured by S11 using single-component \ser fits, for these subsamples. This measurement reflects the relative degree of bulge- or disc-dominance of the galaxies in these samples, as extended exponential discs are fit with shallower light profiles (lower \ser indices), while more centrally-concentrated bulges are fit with more sharply peaked light profiles (higher \ser indices). Consideration of the \ser index distributions enriches the picture of the population painted with axis ratios. While the star-forming population ([12]--[3.4] $\la 0$) is indeed disc-dominated overall, showing \ser indices in the 1-2 range, at high luminosities bulges are somewhat more prominent than at faint luminosities. Bulges are much more prominent for the quiescent population ([12]--[3.4] $\ga 1$), with most of these galaxies showing \ser indices of 3-4 or more; the 3-D shape of these bulges (shown by their axis ratio distribution) is a function of luminosity, mirroring the distinction between fast rotators at low luminosity and slow rotators at high luminosity \citep{emsellem_etal_11}. 

\section{Analysis \& Results} \label{sec:results}

The galaxies which occupy any given location in our M$_{3.4\um}$ - [12]--[3.4] parameter space should be close to identical in terms of their stellar populations. As such, they should have equal intrinsic magnitudes $M_{\lambda}$ in any optical band $\lambda$. The optical luminosity suffers from dust attenuation while the 3.4\um NIR luminosity does not (or, at least, does to a much smaller extent). Since we observe these galaxies at an arbitrary range of viewing angles and this attenuation will be more severe when viewed closer to edge-on, the observed $M_{\lambda} - M_{3.4\um}$ (henceforth noted as [$\lambda$]--[3.4]) colour of these galaxies will vary as a function of inclination. If one makes the assumption that the structures of the galaxies in that bin in stellar populations are drawn from a population of flattened discs viewed from random directions (consistent with observed axis ratio distributions, except for luminous non-star forming galaxies; Fig.\ \ref{fig:btba_duck}), the dependence of [$\lambda$]--[3.4] on inclination should correspond directly to a dependence of [$\lambda$]--[3.4] on {\it axis ratio}. It is this dependence which we seek to characterize in this section.

However, the intrinsic (rather than observed) [$\lambda$]--[3.4] colour of galaxies also varies as a function of $M_{3.4\um}$, both between bins (which does not affect our analysis) and even across an individual bin (which does). Therefore, within each bin we fit the relationship between [$\lambda$]--[3.4] colour and $M_{3.4 \um}$ with a linear relation and remove it. We shear the [$\lambda$]--[3.4] - $M_{3.4 \um}$ distribution around the centre of the $M_{3.4 \um}$ bin, which removes the $M_{3.4 \um}$ dependence while preserving the average [$\lambda$]--[3.4] colour of the bin (this is not, strictly speaking, necessary, as the normalization of colour is arbitrary in our analysis since we are concerned only with the \emph{variation} of colour with axis ratio, but this facilitates more intuitive comparisons between bins). For each bin, we denote the resulting galaxy colours as [$\lambda$]--[3.4]$'$.\footnote{This correction is carried out because this is the proper way to conduct this analysis, but its presence (or absence) has only minimal effect on the results (see \S\ref{subsec:errors}).}

As described in \S\ref{subsec:i_indep}, at this stage we impose a redshift cut within each bin to guard against biases due to sample selection limits in apparent magnitude and radius. On average this removes roughly 50-60 per cent of the galaxies in each bin, but while this does result in somewhat larger statistical errors, as long as the number of galaxies does not drop too low this does not significantly hamper the measurement of the attenuation. 

\begin{figure}
\begin{center}
\includegraphics{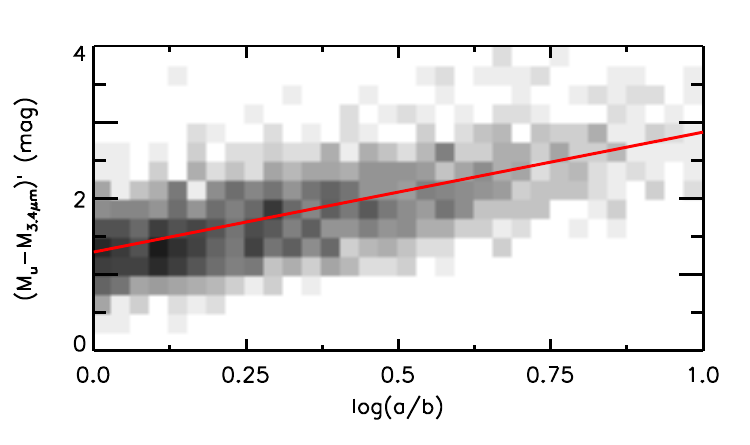}
\caption{Distribution of galaxies as a function of [$u$]-[3.4]$'$ colour and axis ratio for a representative example bin of parameter space located at $-22 \leq M_{3.4\um} \leq -21$ in luminosity and $-1.0 \leq [12]-[3.4] \leq -0.5$. The 2-d histogram shows the galaxy distribution, while the solid line shows the best-fitting linear relation to this distribution. The slope of the line is the attenuation parameter $\gamma$ for this bin. \label{fig:slpfit}}
\end{center}
\end{figure}

Finally, we quantify the inclination dependence of luminosity in any given bin with the commonly-used parametrization (e.g. \citealt{tully_etal_98}, \citealt{maller_etal_09}) $A_{\lambda} = \gamma_{\lambda} \log(a/b)$, where $A_{\lambda}$ is the attenuation relative to face-on in the passband $\lambda$, $a/b$ is the ratio of the semi-major to semi-minor axis (inverse of the axial ratio), and $\gamma_{\lambda}$ is the resulting attenuation amplitude parameter in that band. This is done by fitting a linear relationship to the [$\lambda$]--[3.4]$'$ - log($a/b$) distribution of galaxies -- the slope of this relationship is the amplitude parameter $\gamma_{\lambda}$ of interest.\footnote{The total face-on vs.\ edge-on relative attenuation for a disc galaxy, in mag, is thus $\tld0.85 \gamma$, as the run in log($a/b$) using S11's axis ratios is \tld0.85 for most disc-dominated bins of parameter space.} This procedure is illustrated in Fig.\ \ref{fig:slpfit}, which shows the galaxy distribution and fit relationship in SDSS $g$ band for a representative example bin in our parameter space located at $-22 \leq M_{3.4\um} \leq -21$ in luminosity and $-1.0 \leq [12]-[3.4] \leq -0.5$ -- this represents star-forming disc galaxies broadly similar to the Milky Way.

\begin{figure*}
\begin{center}
\includegraphics{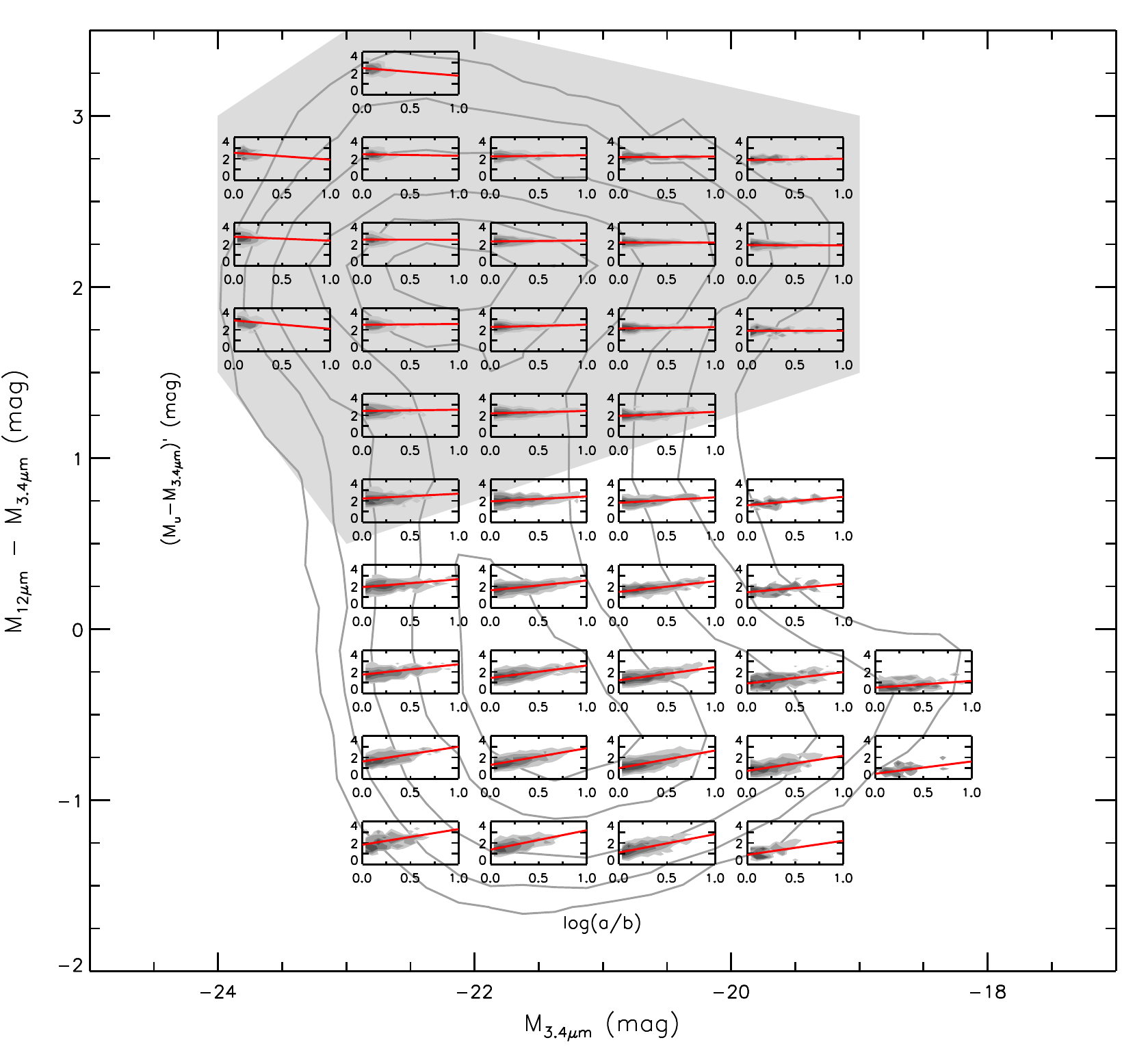}
\caption{Background contours show the distribution of galaxies in [12]--[3.4] vs $M_{3.4\um}$ space. The inset panels show the distribution of [$u$]--[3.4]$'$ colour vs log($a/b$), and the best-fitting line thereof, for galaxies at that location in [12]--[3.4] vs $M_{3.4\um}$ space. The slope of the best-fitting line represents the amplitude of the attenuation at that point in the parameter space. Grey fill shows approximately the region of parameter space occupied by bulge-dominated galaxies. \label{fig:slpfit_duck}}
\end{center}
\end{figure*}

The use of a linear fit to parametrize the relationship between $A_{\lambda}$ and log($a/b$) has been challenged in the past. For one example, \citet{masters_etal_10}, in studying the variation of NIR colours with axis ratio, suggested that a bilinear or quadratic fit may be more appropriate. Indeed, a linear relationship is not required by any physical law, but rather represents the simplest functional form that fits the data to within the scatter. While the true physical form of the relationship may be substantially more complex, the distribution of colours as a function of axis ratio in Figs.\ \ref{fig:slpfit} and \ref{fig:slpfit_duck} show little in the way of systematic deviation from a linear fit that is outside the random or systematic uncertainties. For this reason, we feel comfortable retaining a linear relationship between $A_{\lambda}$ and log($a/b$), especially as this enables a far more direct comparison with most previous work.

It is also important to note that this parametrization depends significantly on the range of axis ratios covered by the sample in the chosen set of input measurements. Since $\gamma_{\lambda}$ is the \emph{slope} of the relationship between $A_{\lambda}$ and $\log(a/b)$, variations in the axis ratio distribution within a bin between different sets of input axis ratio measurements can alter the resulting slopes even for an identical colour distribution. (We explore this issue further in \S\ref{subsec:errors}.) This parametrization is thus in some ways not ideal, but for now we retain it in the interest of comparison with previous work. We may revisit this issue in the future.

We apply this method in bins of $M_{3.4\um}$ luminosity and [12]--[3.4] colour throughout our parameter space. Each bin is 1 magnitude wide in luminosity and 0.5 magnitudes wide in colour, and our parameter space contains sufficient galaxies to populate bins in the area $-24 \leq M_{3.4\um} \leq -18$ in luminosity and $-1.5 \leq [12]-[3.4] \leq 3.5$. (Galaxies can, of course, be found outside these limits, but not in sufficient numbers for reliable analysis.) A schematic outline of our analysis across this parameter space can be found in Fig.\ \ref{fig:slpfit_duck}.

Fig.\ \ref{fig:slpfit_duck} shows both the overall distribution of galaxies within the parameter space (background contours), and the [$u$]--[3.4]$'$ vs. log($a/b$) distributions and attenuation fits for the galaxies in each bin (inset panels). The variation in attenuation across our parameter space can be seen in the varying slopes of the attenuation fits in each bin. For each inset panel of Fig.\ \ref{fig:slpfit_duck}, the y-intercept of the fit represents the face-on [$u$]--[3.4] colour of the galaxies in that bin. It can be seen that this colour varies with $M_{3.4\um}$ luminosity, becoming redder for brighter galaxies. This has been studied a number of times in the past (e.g., \citealt{dejong_96}, \citealt{bell_dejong_01}), and this variation reflects a combination of variations in star formation history, metallicity, and dust content with luminosity.

\begin{figure*}
\begin{center}
\includegraphics{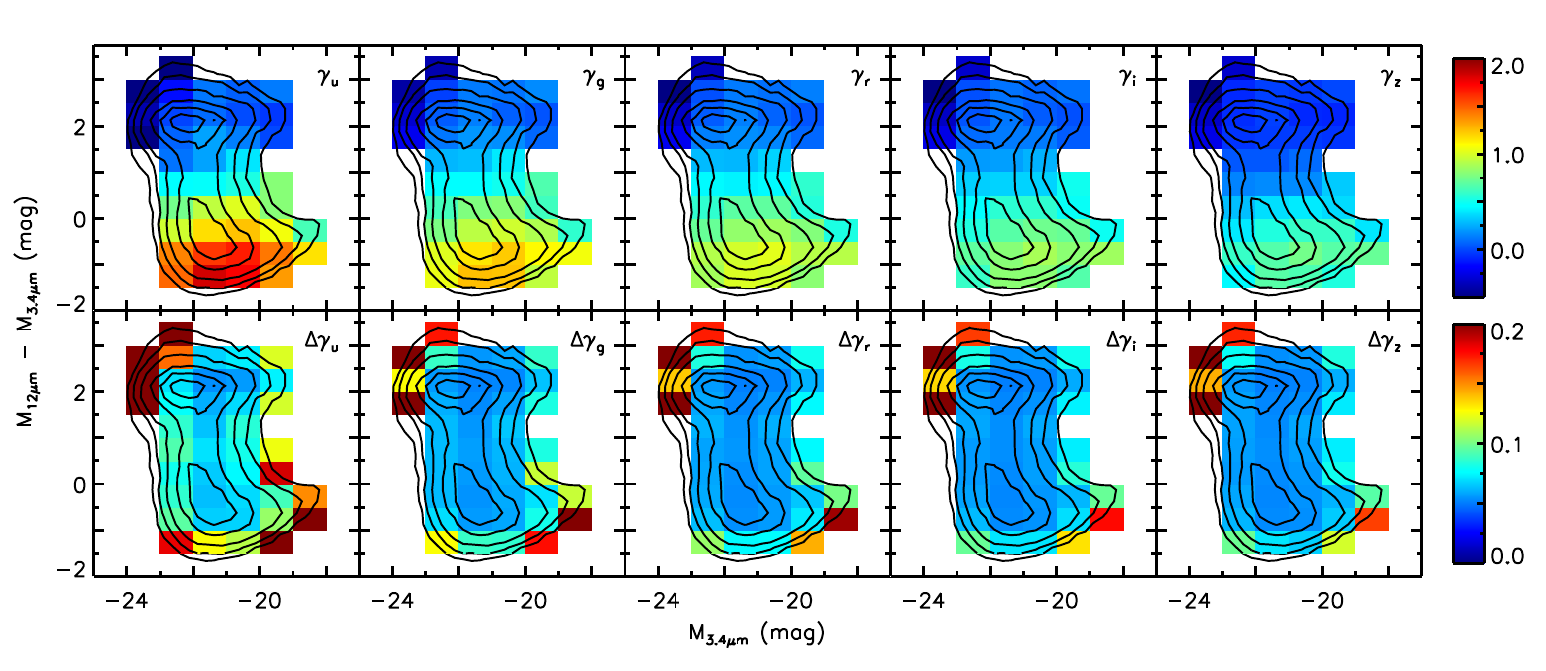}
\caption{Top row: Dust attenuation amplitude parameter $\gamma$ as a function of $M_{3.4\um}$ luminosity and [12]--[3.4] colour throughout our parameter space, for the 5 SDSS bands $u, g, r, i, z$. As in Fig.\ \protect\ref{fig:mstarssfr_ducks}, contours show the distribution of galaxies within this parameter space. Colours show the value of the attenuation parameter $\gamma$ for galaxies within each bin, and the colour scale is the same for all bands. Bottom row: Same as top row, except colours show the uncertainties in the attenuation parameter $\gamma$.\label{fig:gamma_ducks}}
\end{center}
\end{figure*}

\subsection{Attenuation as a function of wavelength and galaxy parameters} \label{subsec:atten}

Our results for all 5 SDSS bands \emph{u, g, r, i, z} are presented graphically in Fig.\ \ref{fig:gamma_ducks} and numerically in appendix A in Tables \ref{tab:atten_u}--\ref{tab:atten_z}. The colour coding represents the size of the attenuation amplitude parameter $\gamma_{\lambda}$, with blue colours representing low values (less or no dependence of [$\lambda$]--[3.4]$'$ colour on axis ratio, and so little or no dust attenuation) and red colours representing high values (strong dependence of [$\lambda$]--[3.4]$'$ colour on axis ratio, and so large amounts of dust attenuation). The colour scale is identical throughout, allowing comparison between the bands.

First, and most obviously, the attenuation is stronger in the shorter wavebands. While the amplitude of the attenuation varies with band as expected, the patterns with galaxy properties within each band are extremely similar. The effects discussed below are easiest to see in the $u$ band in Fig.\ \ref{fig:gamma_ducks} since this band has the strongest attenuation, but these effects apply with very little modification to the other bands as well.

Second, there is a very obvious gradient in attenuation strength with [12]--[3.4] colour. Near the top of the diagram we find quiescent, bulgy galaxies. For many of the most luminous or least negative [12]--[3.4] colour bins, their axis ratio distributions are so narrow (Fig.\ \ref{fig:btba_duck}) that our method does not produce meaningful results, as these are obviously not discs but rather spheroids and any variation in axis ratio is intrinsic rather than due to random inclination. Therefore, the near-zero (or even negative) values of $\gamma_{\lambda}$ for these bins are not physically meaningful. However, many of the less luminous and/or more negative [12]--[3.4] colour bins in this area have sufficiently broad axis ratio distributions to indicate the presence of flattened discs seen from a variety of viewing angles, and for these we can see that they are consistent with little or no dust attenuation, consistent with our interpretation of these systems as quiescent, with little in the way of cold gas or dust content.

However, as one moves to more negative [12]--[3.4] colours, the attenuation increases strongly towards the star-forming main sequence. This area is populated by star-forming disc-dominated galaxies which have notable dust attenuation. Within this region, there are two major parameters that correlate with attenuation -- luminosity and star formation activity.

First, attenuation varies as you move along the star-forming main sequence from faint to bright at a given [12]--[3.4] colour. The dimmest galaxies have the least attenuation, and attenuation increases as you move towards brighter galaxies, similar to the results of previous studies. However, in our case we find that attenuation reaches a maximum for galaxies of intermediate luminosity, and begins to decrease again for the most luminous galaxies. This (seemingly) contradicts many studies which have found a monotonically increasing dependence of attenuation on luminosity (e.g. \citealt{tully_etal_98}, \citealt{maller_etal_09}). However, these studies generally have not probed to high enough luminosities to match our analysis -- for example, Tully et al. are limited to roughly $K \geq -23.25$ and Maller et al. consider only galaxies with $K \geq -22.75$ (corresponding to $M_{3.4\um} \approx -22.0$ and $-21.5$), neither of which pass the turnover region we find by enough of a margin to discover this effect (see discussion in \S\ref{sec:tully_maller}).

Second, as one crosses the star-forming main sequence vertically from the least to the most star-forming galaxies at a given luminosity the attenuation increases, reaching a maximum for the galaxies with the most negative [12]--[3.4] colours at the very bottom edge of the star-forming main sequence. In other words, not only does [12]--[3.4] colour divide star-forming dusty galaxies from quiescent non-dusty ones, but within the star-forming dusty galaxies the dust attenuation is a strong function of specific star formation. Notably, the variation of attenuation with specific star formation rate is comparable in magnitude to, or even stronger than, the variation of attenuation with luminosity.

Finally, in addition to the obvious variations with $M_{3.4\um}$ luminosity and [12]--[3.4] color within a given band and the variations in overall attenuation between bands, there are hints that the location of strongest attenuation shifts slightly fainter along the star-forming main sequence in the longer wavebands. It is difficult to see in Fig. \ref{fig:gamma_ducks}, but perusal of the numerical data in Tables \ref{tab:atten_u}--\ref{tab:atten_z} shows that the strongest attenuation in the $r$, $i$, and $z$ bands is located at $M_{3.4\um} \approx -20.5$ rather than $-21.5$, though the differences are slight (especially in the $r$ band, where they are essentially negligible). 

\subsection{Statistical \& systematic uncertainties and the effects of varying input measurements} \label{subsec:errors}

Errors for the attenuation parameters are presented graphically in Fig.\ \ref{fig:gamma_ducks} and numerically in appendix A in Tables \ref{tab:atten_u}--\ref{tab:atten_z}. Statistical errors are calculated by bootstrap resampling, and are generally of order \tld0.04--0.06 in slope, though they increase near the edges of the populated regions of our parameter space where there are fewer galaxies per bin. They vary somewhat with wavelength, being slightly greater in the shorter wavelength $u$ and $g$ bands and slightly smaller in the longer wavelength $r$, $i$, and $z$ bands, but the difference is small.

Systematic errors are somewhat more difficult to quantify. One potential source of error is the various methodological choices we have made in our analysis. In an attempt to quantify this source of error, we have repeated our analysis while varying or omitting some of these refinements. We have experimented with the effects of omitting the K-corrections, omitting the redshift cut within each bin, increasing or decreasing the threshold for the redshift cut, varying the choice of linear fit algorithm, and omitting the shear correction for the $M_{3.4\um}$~-- [$\lambda$]--[3.4] correlation in each bin. Generally we find that these give variations of approximately the same order as (or slightly smaller than) most of the existing statistical errors, \tld0.03--0.05 in slope, and do not depend strongly on location in the parameter space. They vary slightly with wavelength in the same manner as the statistical errors, but again the difference is not large. Therefore, these are conservatively quantified by taking the statistical errors and adding in quadrature an extra slope uncertainty of 0.05, resulting in the final errors.

\begin{figure}
\begin{center}
\includegraphics{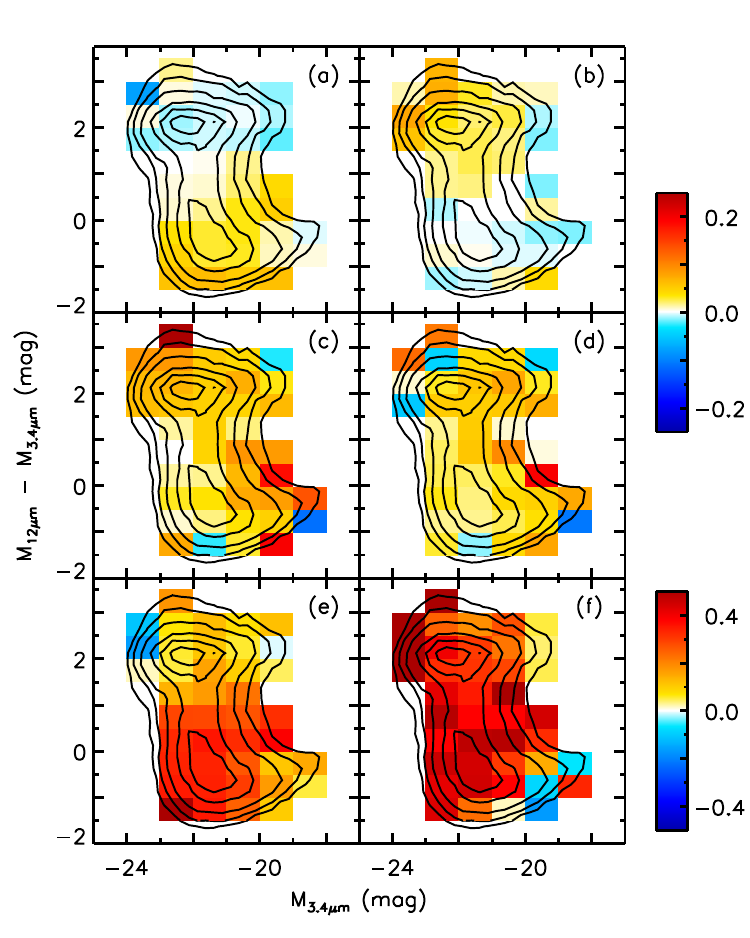}
\caption{Example $g$-band attenuation measurement differences due to differing choices of input parameters, as a function of $M_{3.4\um}$ luminosity and [12]--[3.4] colour. The colour scale shows the difference between the attenuation measured within a bin using a particular alternate input parameter and that measured using the canonical set of parameters. Background contours show the distribution of galaxies in this parameter space.
By panel, the alternate input parameters used are: panel a: SDSS model magnitudes, panel b: SDSS Petrosian magnitudes, panel c: \emph{WISE} profile magnitudes, panel d: \emph{WISE} aperture magnitudes, panel e: SDSS axis ratios, panel f: 2MASS axis ratios. Note that the colour scale in the last two panels covers twice the range as the others. \label{fig:sys_err_ducks}}
\end{center}
\end{figure}

In addition to the statistical and systematic errors, it is also interesting to study the effects of the choice of input measurements used, namely magnitudes and axis ratios. As might be expected, varying these choices significantly alters the measured attenuation parameters. Notably, the alterations due to varying the input measurements are generally significantly larger than those due to the variations in analysis methods mentioned above, and also often show variations with wavelength and across our parameter space.

As discussed in previous sections, we believe that our choices of the SDSS cmodelmags for optical magnitudes, the \emph{WISE} gmags for IR magnitudes, and the S11 axis ratio measurements represent the best available measurements. Therefore, we do not consider the differences presented here to represent any additional error in our work. Rather, we examine the ways in which varying the input data could alter the results in order to help illuminate why we made the choices we have and show the potential consequences of using possibly biased inputs. These effects are illustrated in Fig.\ \ref{fig:sys_err_ducks}.

For optical magnitudes, our other options besides the SDSS cmodelmags were the standard modelmags and the Petrosian magnitudes. Using either of these alternate magnitude measurements significantly alters the measured attenuation values in ways that depend both on location within the parameter space and on passband.

When using the standard modelmags, the most notable deviation within the parameter space is the variation along the star-forming main sequence, with the brightest star-forming galaxies consistently showing an increase in measured attenuation of \tld0.05 in slope relative to the dimmer star-forming galaxies. That is, the use of standard modelmags preferentially boosts the measured attenuation of the brightest star-forming galaxies compared to the results derived with the cmodelmags. In addition to this effect, the overall attenuation of the entire star-forming main sequence relative to that derived from the cmodelmags is enhanced for $u$ and $g$ bands and decreased for $i$ and $z$ bands -- in other words, the attenuation curve is steeper when measured using the standard model mags. This effect increases the average attenuation parameter difference between the $u$ and $z$ bands by roughly \tld0.2. While the exact cause of the size and patterns of variation of these differences is not necessarily apparent without a more detailed analysis, it is notable that since the standard modelmag method fits galaxies with only \emph{either} an exponential or a \dev profile, the bright star-forming galaxies that it assigns higher attenuations to (which contain both strong bulge and disc components) are exactly the ones that the modelmags might be most expected to have trouble with.

Unlike the modelmags, using the Petrosian magnitudes does not cause an increase in variation between different parts of the star-forming main sequence. Rather, the use of the Petrosian magnitudes mostly acts to weaken the divide between the star-forming and quiescent galaxies by increasing the attenuation slope parameters measured for the non-star forming galaxies near the top of our parameter space. This increase is roughly of order \tld0.1 in slope, and varies non-monotonically with magnitude with the $u$ and $z$ bands -- the bands with the poorest S/N for petrosian magnitudes -- being slightly more strongly affected. The star-forming main sequence, on the other hand, while not showing any particular increase in variation within any given band, does again show a steepened attenuation curve, with the measured attenuation decreasing slightly in the $r$ band and more significantly in the $i$ and $z$ bands. This results in an increase of roughly \tld0.05 in the average attenuation difference between the $u$ and $z$ bands, significantly smaller than the similar increase due to the use of the modelmags. Like the variations due to using the modelmags, the exact causes of the details of these variations is not especially apparent without a more detailed analysis, but again, it is notable that the quiescent galaxies (i.e. galaxies with \dev light profiles) which are strongly affected are ones that the Petrosian magnitude measurement method is known to have difficulty measuring.

The other major choice of photometric input measurements we made was to use the 2MASS-derived \emph{WISE} gmags for our NIR photometry, rather than the standard aperture or profile-fit measurements. As noted in \S\ref{subsec:wise} these alternatives are not well-suited to measuring extended sources such as our galaxies, and, as with the alternate SDSS photometric measurements, using them results in significant alterations in measured attenuation.

The use of either the standard aperture or profile-fit photometry results in roughly similar variations in measured attenuation. In both cases, measured attenuation increases for quiescent galaxies by roughly \tld0.05, in a band-independent manner. For star-forming galaxies, the peak of highest attenuation becomes notably broader (which has the effect of also increasing measured attenuation for most star-forming galaxies by \tld 0.05) and shifts roughly 1 mag fainter along the star-forming main sequence. Finally, once more the attenuation curve for star-forming galaxies is affected, in both cases becoming shallower by roughly \tld0.07 in the average attenuation difference between $u$ and $z$ bands.

Finally, and in many ways most critically, we also chose to use S11's axis ratio measurements instead of those from the SDSS or 2MASS. Axis ratio is the choice of input measurement with by far the largest potential effect on output attenuation measurements. As noted previously, this is largely due to the mechanics of this analysis method, since attenuation is expressed as the slope of the colour-axis ratio relationship. This is a notable drawback of this particular choice of parametrization, since it can create or exaggerate differences between studies even when the analysis techniques and other galaxy properties are similar or identical. One could in principle correct for this by scaling to the range of axial ratios present in the sample, but this has not been commonly done in past studies. Therefore, since the variation in axis ratio measurements between different data sets can be quite large, and moreover is often strongly correlated with galaxy properties, the choice of axis ratio measurement is critical.

Illustrating this, using SDSS axis ratios rather than those of S11 results in significantly increased attenuation measurements. This increase is quite large -- up to 0.5 in slope in the $u$ band -- and is by far the strongest among the brighter galaxies along the star-forming main sequence. As with the modelmags, this is likely due to the presence of significant bulge components in these galaxies. These result in much rounder measured axis ratios than would otherwise be expected for populations of randomly-inclined disc galaxies, narrowing the axis ratio distributions within each bin and therefore steepening the slopes of the colour-axis ratio relationships in those bins. Notably, this effect is strongest for the brightest disc galaxies, and is great enough that when using these axis ratios the measured attenuation is also strongest for these brightest galaxies, rather than peaking at intermediate luminosities as we observe. 

Similarly, using 2MASS axis ratios also results in significantly increased attenuation measurements. This increase is even greater than for the SDSS measurements, reaching more than 0.7 in slope in the $u$ band. This effect tends to be slightly larger for the most luminous galaxies, likely reflecting the same influence of bulges as in the SDSS, but overall is much more widespread. This is likely due to the lower resolution of 2MASS data, which (as noted in \S\ref{subsec:ab}) results in unreliable shape measurements for the smaller galaxies which comprise most of our sample. For these small galaxies, seeing effects will tend to result in rounder measured axis ratios regardless of the presence (or absence) of a bulge, further increasing measured attenuations across the board. 

These strong effects emphasize the importance of very careful measurements of axis ratios, sharply highlight the limitations of low resolution, purely isophotal, or single-model measurements. We believe that S11's axis ratio measurements are the best currently available by a significant margin, but further developments in this area would be welcome, especially in longer-wavelength measurements. These effects also illustrate the limitations of this method of parametrizing attenuation, and we may attempt to mitigate this issue in future work.

\section{Comparisons to previous work} \label{sec:tully_maller}

Fig.\ \ref{fig:gamma_ducks} illustrates that attenuation is more significant at shorter passbands than longer ones, shows a strong dependence on our observational proxy for star formation activity ([12]--[3.4]) in that more intensely star forming disc galaxies are more attenuated than their less star forming counterparts, and shows a complex dependence on luminosity, in that attenuation is smaller for both less luminous and more luminous galaxies than the Milky Way (at $M_{3.4\um} \approx -21.5$). The goal of this section is to compare our results to two immediately comparable works -- \citet{tully_etal_98} (hereafter T98) and \citet{maller_etal_09} (hereafter M09). 

Our work, T98's, and M09's use similar methods, in that all works assume that at sufficiently long wavelength galaxy luminosity is effectively independent of inclination. After removing the overall trend in optical--NIR colour vs.\ NIR luminosity, any remaining trends in $M_{\lambda} - K$ colour with inclination are interpreted as dust-induced attenuation, and are presented as the slope of the relation between optical--IR colour and the logarithm of axial ratio. Yet, the three works have significant differences that impede straightforward comparison of attenuation relation slopes: the works are based on different data, parametrize attenuation differently, and use different axial ratio measurements. Given that the intention of each of these works is to provide an inclination correction that can return galaxy magnitudes to their face-on values, the final inclination corrections themselves are a reasonable point of comparison.

As such, to compare our work with that of T98 and M09, we calculate three different inclination corrections for each galaxy in our sample based on the parametrization and results of each work. For our work, we calculate an inclination correction using our SDSS $g$-band attenuation relation slope at the galaxy's given S11 axial ratio, $M_{3.4\um}$ absolute magnitude and [12]--[3.4] colour. For M09, the SDSS $g$-band attenuation relation slope is derived according to the results in their Table 2, parametrized in terms of absolute $K$ magnitude (from 2MASS) and concentration (from SDSS), and the inclination correction is calculated using that slope and SDSS model fit axial ratios. For T98, the attenuation relation slope is parametrized in terms of $B$-band absolute magnitude, which we derive from our SDSS magnitudes using {\sc P{\'e}gase} (Fioc \& Rocca-Volmerange 1997) stellar population synthesis models. The inclination correction in $B$ band is then calculated using that slope and S11's axial ratios, and translated back into SDSS $g$ band using the attenuation curve of \citet{calzetti_00}. This yields a predicted inclination correction in SDSS $g$ band for each galaxy using each of the three works' methods.

\begin{figure}
\begin{center}
\includegraphics{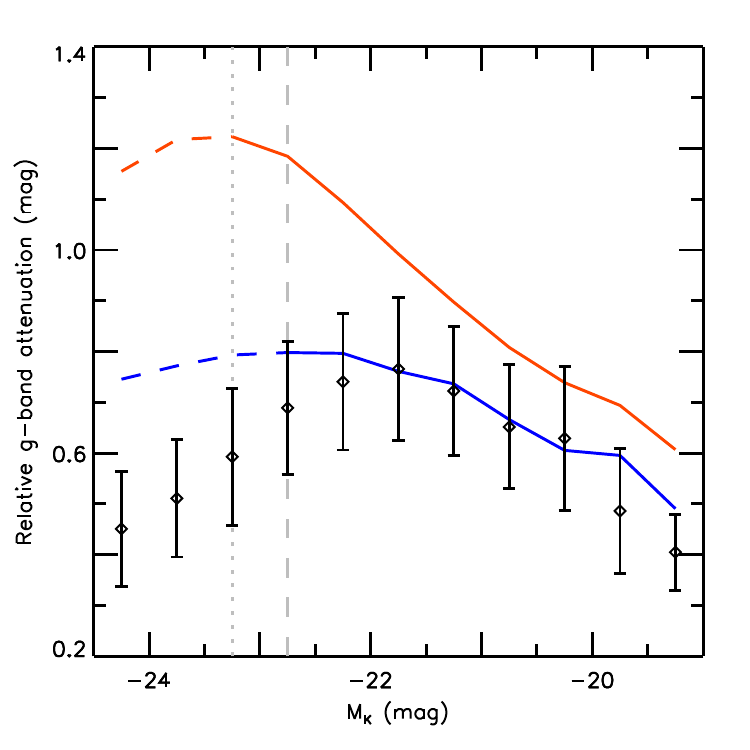}
\caption{Predicted edge-on relative $g$-band attenuation of main sequence star-forming disc galaxies as a function of $K$-band luminosity. Black data points are our work, while red and blue lines are predictions based on the results of T98 and M09 respectively. The vertical dotted and dashed lines represents the approximate magnitude limits of the data used in those two works, and the switch to dashed lines beyond these limits shows that those attenuation estimates are extrapolations. \label{fig:devour_tully_maller}}
\end{center}
\end{figure}

In Fig.\ \ref{fig:devour_tully_maller}, we compare the predicted inclination correction for our work with those of T98 and M09, in this case as a function of $K$-band absolute magnitude. Since T98 and M09 attempted to select disc-only samples, we restrict this comparison to include only galaxies which lie along the star-forming main sequence as defined by our $M_{3.4\um}$ and [12]--[3.4] selection. In each case, the presented inclination correction is the average correction for the 20\% most inclined members of the population at that magnitude, and the error bars are the scatter among that group.

We can see that at lower luminosities, all three methods predict a similar pattern of increasing inclination corrections with luminosity. M09's predictions match ours very closely, while T98 predicts stronger attenuation by up to 0.2 mag (the reasons for this discrepancy will be discussed below). At higher luminosities, however, both T98 and M09 predict much stronger attenuation, as their predicted corrections do not significantly decrease as ours do. Both the overall discrepancy with T98 and the dramatic differences between our work and those of T98 and M09 at high luminosities appear concerning, and we investigate (and largely resolve) them below. 
 
\subsection{Comparison with Tully et al.} \label{subsec:tully}

T98 measured inclination dependence in optical--NIR colours as a function of absolute magnitudes in BRI using a very similar methodology to ours. Their sample is much more limited, containing only 100 galaxies spread between two clusters, but both are relatively nearby and the quality of the data is generally good. Given the similarity of our methods, the offsets in inferred attenuation in Fig.\ \ref{fig:devour_tully_maller} seem surprising. 

In order to investigate this further, we choose to repeat the analysis of T98 using our joint \emph{WISE}-2MASS-SDSS dataset. We convert the SDSS $ugriz$ absolute magnitudes into $BRI$ using the {\sc P{\'e}gase} stellar population synthesis models and convert our attenuations into $BRI$ using the attenuation curves of \citet{calzetti_00}, and bin the sample in $K$-band luminosity only rather than in both luminosity and [12]--[3.4] colour.

\begin{figure*}
\begin{center}
\includegraphics{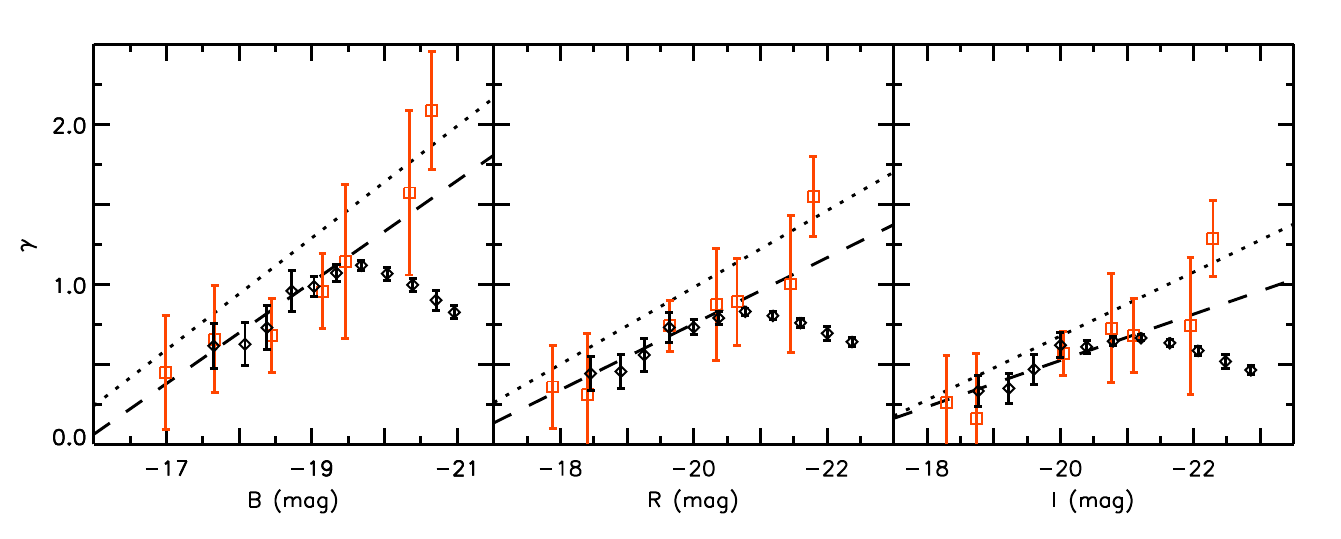}
\caption{The strength of the attenuation parameter $\gamma$ as a function of absolute $B$, $R$ and $I$ magnitudes. Diamonds are the values for star-forming galaxies from our sample, while red squares are data from T98. Dotted lines are T98's joint fits to their and \protect\citet{giovanelli_etal_95}'s data, while dashed lines are fits to T98's data alone (neglecting the highest luminosity data point). \label{fig:BRI_devour_tully}}
\end{center}
\end{figure*}

The results are shown in Fig.\ \ref{fig:BRI_devour_tully}. The diamonds show our estimates of the slope of the attenuation relation $\gamma$ as a function of luminosity in different bins of absolute $B$ (left panel), $R$ (middle panel), or $I$ (right panel) magnitude. Meanwhile, the squares are T98's estimates, and the dotted lines are the fits that T98 adopt for their final attenuation prescriptions (that is, T98's equations 3, 4, and 5). The dashed lines will be explained shortly.

As can be seen, our data are broadly consistent with T98's at low and intermediate luminosities, but depart from T98's results at the high end. Their highest luminosity data point, in particular, is very inconsistent with our data. Additionally, the final fits that T98 adopt are well above our (and their!) results. However, the agreement of their actual data points with ours at lower luminosities is very good. Accordingly, there are two issues to understand: the discrepancy between the presented fits (dotted lines) and the data (both ours and theirs), and the strong attenuation of their highest luminosity subsample.

T98's fits are a joint fit to both their data and data taken from \citet{giovanelli_etal_95}, and it is the influence of both that data and T98's highest luminosity data point that combine to pull their final fits well above our results. It is apparent from a visual inspection of T98's data and fits in Fig. \ref{fig:BRI_devour_tully} that their overall fits are inconsistent with their lower luminosity data points. Indeed, fits simply to T98's data alone, discarding the highest luminosity data point (represented by the dashed lines in Fig.\ \ref{fig:BRI_devour_tully}), match our results much more closely.

\begin{figure*}
\begin{center}
\includegraphics{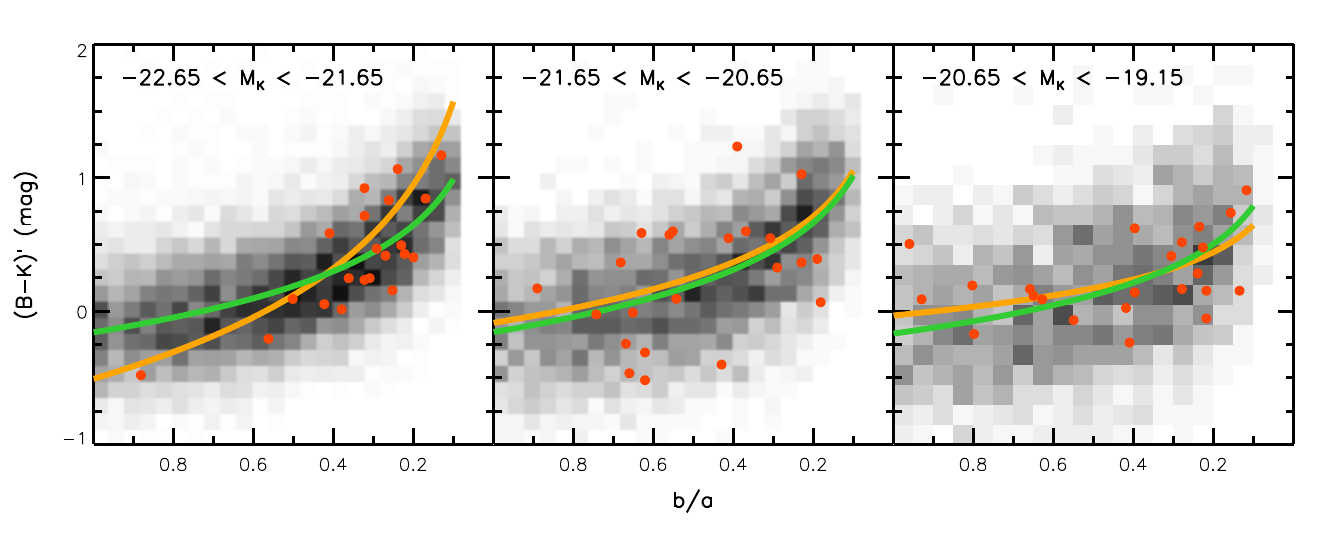}
\caption{Galaxy $B-K$ colour residuals as a function of axis ratio for star-forming galaxies from our sample, divided into bins of $K$-band luminosity. Green lines are the best-fit relation between log($a/b$) and colour for our sample. Orange lines and red circles are T98's fits and data, respectively.
\label{fig:BK_ba_devour_tully}}
\end{center}
\end{figure*}

Fig.\ \ref{fig:BK_ba_devour_tully} shows why we believe that the highest luminosity data point is erroneous (this figure is an analogue of T98's figure 2). Each panel shows, for one bin in $K$ luminosity, the $B-K$ colour residuals (after the removal of the overall trend in $B-K$ colour with $K$) as a function of axis ratio. The leftmost panel shows this relation for galaxies on the star-forming main sequence with $M_{K} < -21.65$, matching T98's highest luminosity bin. The 2-d histogram shows the distribution of galaxies from our sample, while the red circles are the galaxies from T98's sample. As can be seen, in this luminosity bin T98's sample has a very skewed distribution of axis ratios and contains {\it only one} galaxy more face-on than $b/a=0.6$. Additionally, that one galaxy can be seen to lie (somewhat unfortunately) significantly below the average for its inclination. This causes T98's fit for $\gamma$ in this bin (orange line) to be highly inconsistent with the overall galaxy population, suggesting a far higher attenuation than is actually implied by our more comprehensive sample (green line).

Visually T98's relation may seem to be a better fit to the highly inclined end of the sample. However, we find no improvement in the quality of the fit by adopting a steeper slope at the high end, even if low inclination galaxies are excluded. This visual impression is caused by the fact that the distribution of $B-K$ colour for highly inclined galaxies is significantly non-gaussian. This non-gaussian distribution is in part explained by the [12]--[3.4] dependence in the attenuation explored in Fig.\ \ref{fig:gamma_ducks}; our preliminary explorations indicate that dependence of attenuation on morphological and structural parameters (e.g., bulge prominence) also contribute (Devour \& Bell, in prep.).

The centre and right panels of Fig.\ \ref{fig:BK_ba_devour_tully} expand this comparison to include the luminosity bins $-21.65 < M_{K} < -20.65$ and $-20.65 < M_{K} < -19.15$, matching T98's next two bins. (T98's lowest luminosity bin, $-19.15 < M_{K}$, is very sparsely populated and shows almost no dependence of colour on axis ratio, so we do not include it.) In these cases there are no large skews in the axis ratio distributions, so our fits are much closer to T98's, and indeed the slopes match quite well.

To conclude, the discrepancy between our and T98's fit results is somewhat illusory -- the data are consistent with the data (Fig.\ \ref{fig:BK_ba_devour_tully}). The differences in our fit results are driven by two factors: T98 lacks a representative sample of face on luminous galaxies, driving their attenuation estimates to artificially high values, and that T98 (by hand) modifies their attenuation results to more closely mirror the results of \citet{giovanelli_etal_95}, degrading the match between their `fit' results and their (and our) data. We conclude that our results are both consistent with T98's data, and, owing to the larger and more representative sampling from SDSS/2MASS/\emph{WISE}, are a more faithful representation of the actual trends in attenuation with galaxy properties than T98's analysis could probe. 

\subsection{Comparison with Maller et al.} \label{subsec:maller}

In Fig.\ \ref{fig:devour_tully_maller} it was shown that when both our methods and M09's are parametrized solely in terms of $K$-band magnitude, our results agree for galaxies less luminous than the upper limit of M09's sample, but disagree for brighter galaxies. In this section, we repeat M09's analysis to explore the origin of this seeming discrepancy. 

M09 begin with a sample drawn from the NYU-VAGC, an optical catalog based on the SDSS which fits each galaxy with a \ser light profile to measure its \ser index $n_{s}$, half-light radius $r_{50}$, and axis ratio $b/a$. This sample is cross-matched with 2MASS for NIR magnitudes, and subjected to $r$-band absolute magnitude limits and a minimum $r_{50}$ size limit. They then use a combination of cuts on \ser index and axis ratio ($n_{S} \leq 3.0$ or $b/a \leq 0.55$) to isolate disc galaxies. It is important to note that the NYU-VAGC sample is volume limited, and that this dramatically limits the number (and weight in the fits) of galaxies more luminous than $M_K<-22.75$.  

We note that the use of optical half-light radii and \ser indices is potentially troublesome -- both are likely to be affected by dust attenuation \citep{pastrav_etal_13}, and may lead to biased fit results. We choose to simply mirror their fitting methodology for the present moment, deferring a full discussion of the effects of attenuation on galaxy structures to a future work (Devour \& Bell, in prep.). 

M09 parametrize their results in terms of not only absolute $K$-band magnitude but also in terms of \ser index, and they fit the attenuation by a function of the form $\gamma = \alpha_{0} + \alpha_{K}(M_{K} + 20) + \alpha_{n}n_{s}$ -- in other words, a plane in the $M_{K}$-$n_{s}$ parameter space. This parametrization is the only major mechanical difference between M09's analysis and ours, since we do not assume a functional form for the variation of attenuation across our parameter space.

\begin{figure*}
\begin{center}
\includegraphics{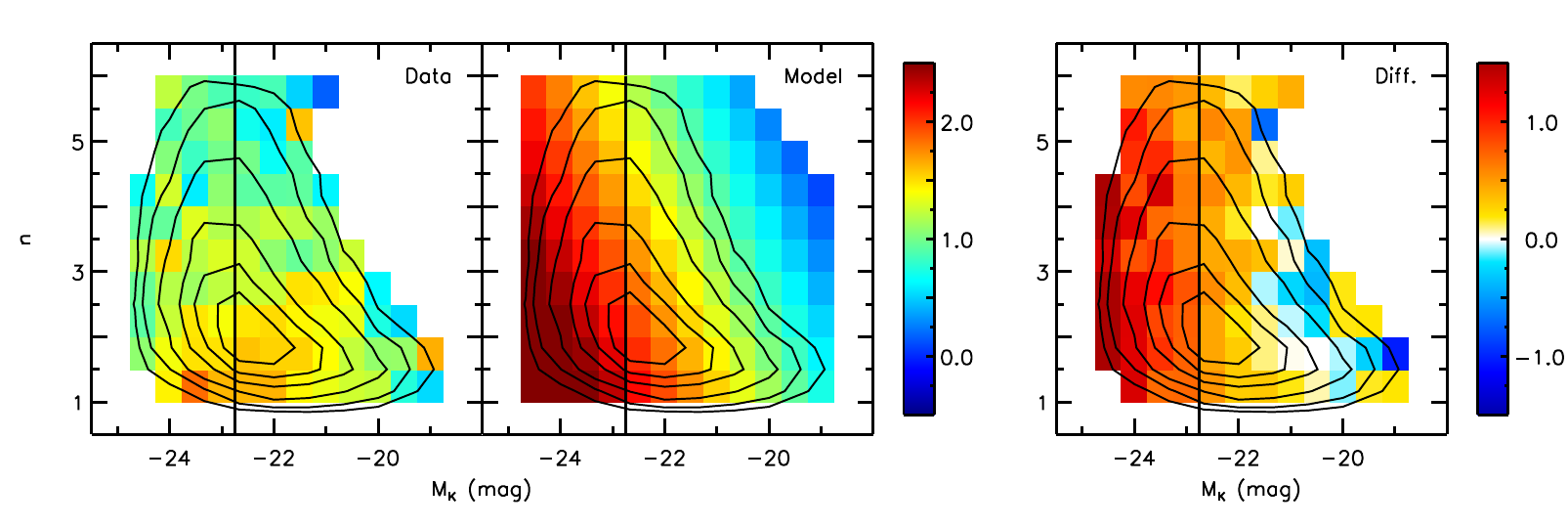}
\caption{Dust attenuation amplitude parameter $\gamma$ as a function of $K$-band luminosity and \ser index, in the $g$ band. Contours show the distribution of galaxies within this parameter space. The dark vertical line shows the magnitude limit of M09's sample. Left panel: measured attenuation parameter using our method. Centre panel: predicted attenuation parameter using the results of M09. Note that, due to the magnitude limit of their sample, all values to the left of the dark vertical line represent an extrapolation. Right panel: Difference of our measured results (left panel) and M09's predicted results (centre panel).
\label{fig:devour_maller}}
\end{center}
\end{figure*}

A comparison between our methodology and M09's can be seen in Fig.\ \ref{fig:devour_maller}. In the left panel, we redo our analysis in exactly the same way as in \S\ref{sec:results}, but following M09's parametrization and input measurements rather than our own. (Namely, parametrizing the attenuation in terms of absolute $K$ magnitude and $r$-band \ser index rather than $M_{3.4\um}$ luminosity and [12]--[3.4] colour, using non-K-corrected\footnote{It is unclear whether or not M09 K-corrected their data. As they do not mention K-corrections at all in their work, we must assume they did not, but regardless this does not significantly affect the results.} SDSS modelmags rather than K-corrected SDSS cmodelmags, and using SDSS $r$-band axis ratios rather than those of S11.) The centre panel shows the predicted strength of attenuation across the same parameter space derived from M09's fit, and the rightmost panel shows the difference between the two.

Overall, both our data and the results produced from it by our method are in substantial agreement with M09 for fainter and intermediate-luminosity galaxies with $M_K>-22.75$. The disagreement seen in Fig.\ \ref{fig:devour_tully_maller} for brighter galaxies is a discrepancy with M09's extrapolation; we argue that their choice of a parameteric plane fit in the $M_K$-$n_s$ plane is not well-motivated by the data, and leads to unfortunate behaviour towards the edges of parameter space, particularly when extrapolated beyond their magnitude limit. We do note that we consider the use of $r$-band \ser index as a controlling parameter for M09's analysis to be a concern and a potential bias, leading us to instead frame our sample selection in terms of attenuation-insensitive near- and mid-IR luminosities (Fig.\ \ref{fig:gamma_ducks}).

\section{Discussion} \label{sec:discussion}

\subsection{Qualitative Comparison with Previous works} \label{subsec:prev}

In addition to the detailed comparisons presented in the previous section, it is instructive to compare our results in a briefer, qualitative sense with the conclusions of a few other works in this area. There have been a wide variety of approaches taken to the study of dust attenuation involving different methods and parametrizations, not all of which are as directly analogous to our own work as those of T98 or M09 but which are still worth mentioning.

\citet{wild_etal_11} study dust attenuation by comparing the SEDs of pairs of galaxies matched in the observed properties of metallicity, specific star formation, axis ratio, and redshift. Their analysis is significantly different than ours, with axis ratio as an explicit parameter rather than our method, and their results are presented almost exclusively in terms of the shapes of attenuation curves, rendering a direct comparison difficult. Nevertheless, it is notable that they find the ratio of continuum to line opacity (representing the strength of the attenuation due to dust in the diffuse ISM) to vary strongly with specific star formation, in accord with our results. They do not directly include galaxy luminosity as a parameter, however they do parametrize on surface brightness, finding that higher surface brightness galaxies have stronger attenuation, but that the variation with surface brightness is not as strong as that with specific star formation.

\citet{driver_etal_07} derive dust/inclination corrections for disc galaxies in the Millennium Galaxy Catalog by matching the galaxy luminosity function at differing inclinations. Their iterative method relies on the fact that the turnover of the luminosity function should occur at the same magnitude regardless of inclination, and uses this to derive general inclination corrections independently for both the disc and bulge components. This method avoids the need to select identical subsamples as we do, but is limited to providing only a single general correction for the entire galaxy population rather than being able to quantify variation in attenuation among that population. They conclude that the average relative attenuation in the $B$ band from face-on to edge-on is approximately 0.9 mag for discs and 1.8 mag for bulges. Comparatively, the nearest equivalent band in our work is SDSS $g$, where we observe a maximum total relative attenuation of approximately 1.0 mag. Their disc attenuations are in general agreement with our results, noting that these are relatively disc-dominated galaxies. However, since these are disc-dominated galaxies it is unclear how their bulge attenuations relate to our results, as the relative contribution of bulge attenuation to the total attenuation will depend on factors such as the distribution of B/T ratios and the geometry of bulges relative to discs.

\citet{masters_etal_10} studied dust attenuation in the SDSS in a very similar manner to our work, parametrizing attenuation in the same way and binning according to galaxy properties to measure the variation in attenuation. In their case, their longer wavelength comparison colour is SDSS $z$, rather than our 3.4\um, and the physical properties they use to parametrize the variation in attenuation are $r$-band luminosity and more `bulgy' or `discy' morphology as measured by the $r$-band SDSS $f_{DeV}$ parameter. Though their methodology is different from that of M09, they also find that less bulge-dominated galaxies have stronger attenuation, though the strength of this variation is somewhat weaker than the variation with luminosity. Given that we find the variation with luminosity to be weaker than that with specific star formation, this would argue that the variation with morphology should also be weaker than that with specific star formation; it will be interesting to see whether this is borne out by our future studies. It is also notable that, unlike older works, they also show a peak in attenuation at intermediate luminosities, in broad agreement with our work. 

\citet{cho_park_09} also study dust attenuation using very similar methodology to our work, again parametrizing attenuation similarly and binning according to galaxy properties. Relative to our work, they use 2MASS $K$ band as their longer wavelength comparison, and parametrize the attenuation by dividing their galaxies according to their $K$-band luminosity and SDSS $i$-band concentration index. They conclude that attenuation is strongest for moderate concentrations and weaker for both the most and least concentrated galaxies, and as with other studies that attenuation is strongest for the most luminous galaxies. However, it is notable that the relationship between attenuation and luminosity noticeably flattens for the most luminous galaxies in their sample, which would be expected in light of our results that attenuation should decrease again for the very brightest galaxies. Another notable aspect is their experimentation with $u-r$ colour (with an attempted correction to the intrinsic, face-on values) as an additional parametrization for dust attenuation. Intrinsic optical colour is related to star formation activity, so based on our work one might expect bluer colours to correspond to greater attenuation. However, they find the opposite, that redder colours correspond to stronger attenuation. This is likely related to the fact that optical colour is also a dust-affected measurement, and so untangling these competing influences on colour is not necessarily straightforward.

\subsection{Optical depth scaling relation predictions \& the importance of star-dust geometry} \label{subsec:scaling}

Our methodology is purely observational, which, as noted previously, has the advantage of not relying on any assumed models. Nonetheless, it is instructive to attempt to reproduce the observed trends using physically motivated tools such as scaling relations. These are useful to gain a quick understanding of the physical principles underlying a set of observations, and it is illustrative to see in what ways these relatively straightforward calculations match our observations, and more interestingly, where they fail to capture the full range of factors that lead to our results.

For this analysis, we use scaling relations to calculate model average dust optical depths for the galaxies in our sample, and then ask how these combine with star-dust geometry to give dust attenuation and how they might be expected to vary across our parameter space.

In the simplest model, the intrinsic optical depth of dust in a galaxy depends on the dust surface density as
\begin{equation}
\tau_{i} = \kappa \Sigma_{d},
\end{equation}
where $\kappa$ is the dust opacity. The dust density is not a measured property of our sample, but it can be derived using scaling relations based on our catalog properties. One could assume that dust density scales with metal column density, which in turn scales with gas column and metallicity, and so
\begin{equation}
\tau_{i} = \kappa f Z \Sigma_{g},
\end{equation}
where $Z$ is the metallicity, $\Sigma_{g}$ is the gas surface density, and $f$ is the fraction of metals in the form of dust. Finally, the gas surface density could be estimated from the star formation rate using the Kennicutt-Schmidt relation \citep{kennicutt_98}: 
\begin{equation}
\tau_{i} = \kappa f Z \left(\frac{\text{SFR}}{\pi r^2}\right)^{1/1.4}\hspace{-1.9em},
\end{equation}
where $\text{SFR}/\pi r^2$ is the star formation rate surface density.

To calculate the metallicity we use the stellar M/L vs.\ [3.4]--[4.5] colour relation from \citet{cluver_etal_14} along with the $M_{3.4\um}$ luminosity to estimate stellar mass (following \S\ref{subsec:i_indep}), combined with the mass-metallicity relation of \citet{tremonti_etal_04}. The star formation rate is derived from the \emph{WISE} 12\um -star formation rate relation of \citet{wen_etal_14}, and this is combined with the SDSS $r$-band petrosian half-light radii to calculate star formation rate surface densities. The metal-dust fraction is estimated at 0.2 in \citet{bell_03} and at 0.5 in \citet{draine_li_07}; we use the latter value of $f=0.5$. The dust opacity (assuming $u$ band) is taken from \citet{li_draine_01}. Thus, using these scaling relations we can construct a model intrinsic (face-on) dust optical depth for each galaxy in our sample.

This is, by necessity, a relatively rough calculation. For example, $f=0.5$ is uncertain and may vary from galaxy to galaxy \citep{draine_li_07,draine_etal_07}, and the $M_{*}(3.4\um, 4.5\um)$, $Z(M_*)$, and $\text{SFR}(12\um)$ scaling relations all have notable scatter. Additionally, the radius used is an optical one, and there is no guarantee that the optical radius and star-forming radius actually correspond. More importantly, this neglects all geometric effects such as dust clumping, relative stellar and dust scale lengths/heights, bulge effects, etc. As such, we do not expect the quantitative size of the predicted optical depths to necessarily match the true physical values. However, this still provides a useful qualitative estimate of the intrinsic dust optical depth, especially since for comparison with our results we are most concerned with the \emph{variation} across our parameter space (which should be more accurately predicted by this model) rather than the absolute normalization.

\begin{figure}
\begin{center}
\includegraphics{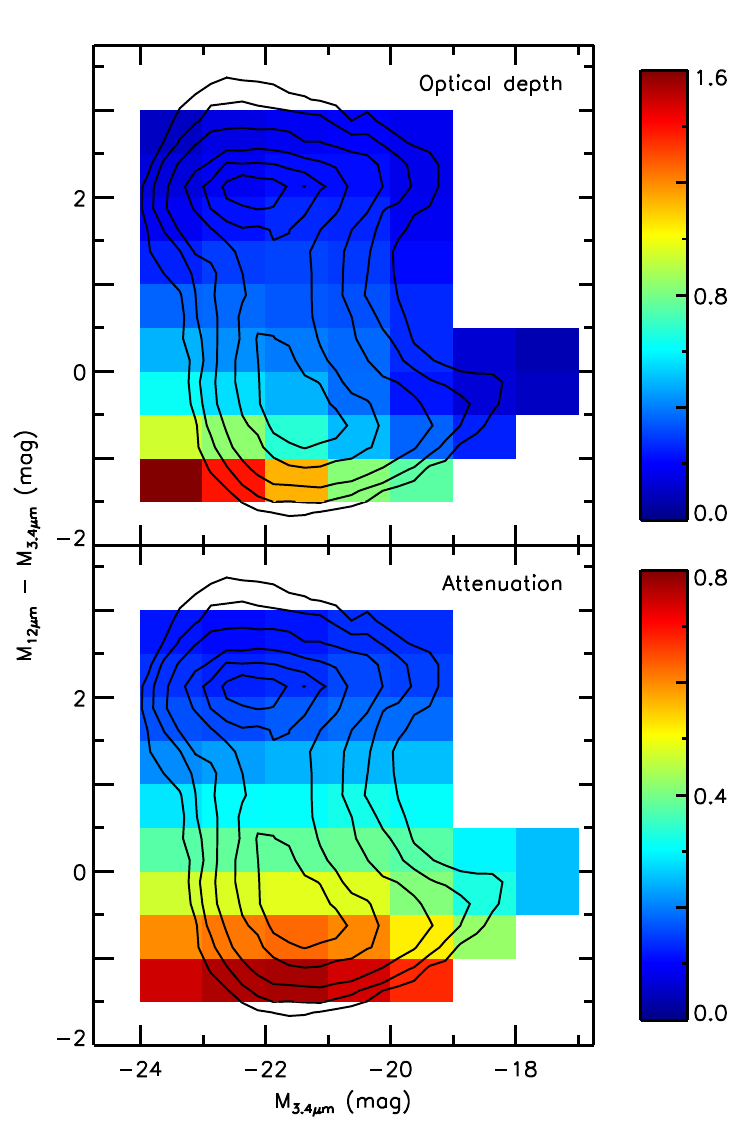}
\caption{Top panel: predicted $u$ band dust optical depth as a function of $M_{3.4\um}$ luminosity and [12]--[3.4] colour, derived from scaling relations based on \emph{WISE} luminosities and SDSS radii. The colour scale shows the intrinsic dust optical depth $\tau$. Contours show the distribution of galaxies within this parameter space. Bottom panel: same as top panel, except the color scale shows the predicted relative face-on-to-edge-on attenuation, derived from the same scaling relations and \protect\citet{tuffs_etal_04}'s disc attenuation models. The colour scale shows the attenuation in magnitudes.} \label{fig:scaling_duck}
\end{center}
\end{figure}

The basic results of this calculation are presented in the top panel of Fig.\ \ref{fig:scaling_duck}, which shows the the average model dust optical depth across our parameter space presented in the same manner as Fig.\ \ref{fig:gamma_ducks}. We can see that the basic patterns apparent in the model calculation are similar to those found in the data. Like our data, the model optical depths show a strong dependence on specific star formation as parametrized by [12]--[3.4] colour, and likewise the models show a significant dependence on stellar mass as parametrized by $M_{3.4\um}$ luminosity. Though this is an intentionally simplified model, this qualitative match is an encouraging sign that the parametrization we use does indeed correlate with variations in dust content, and helps inform intuition about the physical processes that lead to those variations. Moreover, the basic patterns we observe in this model are also seen in more complex ones. For example, \citet{fontanot_etal_09} predict, using a combination of semi-analytic models and radiative transfer simulations, that attenuation in disc galaxies should vary steeply with specific star formation rate as well as with stellar mass (see, e.g., their figure 4), matching our simplified model. 

This model is useful, but it is important to remember that the basic model calculation gives only the (average) \emph{intrinsic, face-on optical depth} -- the total optical depth due to dust, through the entire thickness of the galaxy, as seen by an observer viewing the galaxy face-on. This is not the quantity we infer from our observations: we measure attenuation rather than optical depth, and not only that, we measure the \emph{relative difference} in attenuation between the edge-on and face-on orientations rather than just the attenuation at one particular orientation. Thus, these measurements are not only sensitive to optical depth and star-dust geometry along a certain line of sight, but additionally depend on how the optical depth and geometry vary with viewing angle. Since galaxies contain multiple stellar and dust components with various characteristic shapes and scale lengths, the relationship between attenuation and optical depth can be complex as various components are more or less screened and become optically thick at different inclinations. Thus, the simple face-on optical depth does not necessarily correspond to the full relative attenuation between face-on and edge-on orientations. In order to compare more fully with our observations, we must convert our optical depth model into a relative attenuation model.

This is illustrated in the bottom panel of Fig.\ \ref{fig:scaling_duck}, which shows the results of using our scaling relation optical depths to construct a simple model for the relative face-on to edge-on attenuation. This construction makes use of the model dust attenuations of \citet{tuffs_etal_04} in modified form. \citeauthor{tuffs_etal_04}'s models do not include any variation in galaxy morphology, and their model galaxy contains only a relatively small bulge deeply embedded in a dust disc, leading to very large relative attenuation of bulge light. Our parameter space contains a wide range of galaxy morphologies, and for our purposes we assume that, by the time a galaxy has a bright enough bulge for the attenuation of bulge light to make a significant contribution to the overall attenuation, the bulge is physically large enough to no longer be deeply embedded in the dust disc, in contrast with \citeauthor{tuffs_etal_04}'s assumptions. Therefore, we approximate the overall relative attenuation with disc attenuations only, dividing using the galaxy B/T ratios from S11 and treating the bulge attenuation as being unchanged with inclination. While this is obviously an over-simplification, it is important to remember that this model is intended as a simple, qualitative approximation rather than a fully rigorous physical model. This also correctly treats the largely or entirely bulge-dominated quiescent galaxies as having attenuation largely unaffected by viewing angle. The attenuation for each galaxy is then calculated twice, once assuming a face-on orientation and once assuming an edge-on orientation, and the difference of the two attenuation models then represents the relative face-on to edge-on attenuation.\footnote{The corresponding attenuation parameter $\gamma$ for disc galaxies would then be $(1/0.85)$ times the relative attenuation (see footnote 5) -- we report the relative attenuation here as it is the more physically meaningful property.}

As can be seen, this presents a quite different picture than the basic optical depth model. Most notably, the variation with luminosity is no longer sharply peaked at the highest luminosities, but rather is much more gradual and peaks at intermediate luminosities, matching the patterns we observe in our data. The normalization is significantly different, but this is not unexpected given the simplicity of this qualitative model, and overall the pattern of variations in predicted attenuation with [12]--[3.4] colour and $M_{3.4\um}$ luminosity match our results quite closely throughout the parameter space. More broadly, while this particular relative attenuation model is only intended as a rough qualitative approximation, it illustrates the very strong effects that star-dust geometry and its variation with viewing angle have on attenuation -- the qualitative variation in \emph{observed} relative attenuation with galaxy properties is by no means necessarily the same as the variation in \emph{intrinsic} dust optical depth with those same properties.

\section{Conclusions} \label{sec:conclusions}

Dust attenuation affects nearly every observable property of gas-rich star forming galaxies in the ultraviolet and optical. Because of the strength of the effects of attenuation and ubiquity of dust, statistical study of the attenuation as a function of galaxy parameters has proven challenging -- samples need to be selected and characterized in dust-attenuation insensitive bands, such as the near- or mid-IR, or {\sc Hi}. 

With the advent of large, reasonably deep all-sky near-IR datasets (\emph{WISE} and 2MASS), samples of galaxies detected in the near-IR -- with spectroscopic redshifts in the optical and complementary optical luminosities -- have become available for study. We have used such samples to systematically study the relative dust attenuation in disc galaxies and its variation with NIR luminosity and MIR-NIR colour. Our overall sample, inclination measurements, and photometric measurements are carefully selected to be as free from biases with inclination as possible. We choose to select and classify galaxies using near- and mid-IR measurements where possible (with the only exception being the relatively robust optical axis ratio from S11) in order to avoid any possible dust-induced confusion (and  attendant biases). By selecting our samples in these ways and by using solely observational metrics for our analyses rather than derived physical quantities, we avoid the need for uncertain and model-dependent corrections to our analyses. 

With this data set and classifications, we divide our sample into bins of galaxies which are intrinsically similar in $M_{3.4\um}$ luminosity and [12]--[3.4] colour and measure the relative attenuation between face-on and edge-on galaxies in each bin, thus quantifying the variation in relative attenuation across this parameter space. Our main results are as follows.

The strength of the relative attenuation is very strongly dependent on [12]--[3.4] colour (a proxy for specific SFR, or SFR per unit stellar mass). Galaxies with [12]--[3.4] colours greater than 0.5 show almost no relative attenuation, while below this value attenuation increases very rapidly with decreasing colour (increasing specific SFR) as you move across the star forming main sequence. At greatest, this can represent variation in face-on vs.\ edge-on attenuation of up to 1.1 mag (in the $u$ band) over a span of 2 mag in colour (or roughly an order of magnitude in specific SFR). We conclude that relative dust attenuation is a strong function of specific star formation rate.

The strength of the relative attenuation is also strongly dependent on $M_{3.4\um}$ luminosity (a proxy for stellar mass). However, the variation in relative attenuation with $M_{3.4\um}$ luminosity over the length of the star-forming main sequence (roughly an order of magnitude in stellar mass), while still significant, is generally slightly smaller than the variation with [12]--[3.4] colour. Additionally, the variation in attenuation is not monotonic with luminosity, but rather increases with luminosity to a maximum near $M_{3.4\um} \approx -21.5$ (corresponding to $M_{*} \approx 3\times 10^{10} M_{\sun}$) and then decreases again for more luminous galaxies. We conclude that dust attenuation is also a strong function of stellar mass, though not quite as strong as specific star formation.

Relative to previous results, our result that the strength of the relative attenuation decreases for the most luminous galaxies was somewhat unexpected. Comparing our results in detail with those of T98 and M09, we learn that while our final fit results differ, all the actual datasets match well over the ranges where they overlap. The primary differences between the results come from our probing to brighter luminosities, where we could clearly discern and characterize the downturn in attenuation at bright luminosities where many previous works could or did not. In particular, T98's fits were affected by a somewhat unfortunate unrepresentative small sample at high luminosities, while M09's choice to fit the run of attenuation with luminosity and \ser index with a plane leads to unrealistically high attenuation at high luminosities where their data was sparse.

Simple dust density scaling relations reproduce the strong correlations between dust content, specific star formation rate, and stellar mass, in broad accord with our observational inferences. Making more detailed comparisons with the variation in attenuation across our parameter space, however, requires the use of more complex models that attempt to model the relative face-on to edge-on attenuation. In particular, models which account for differences in star-dust geometry reproduce the non-monotonic variation of attenuation with $M_{3.4\um}$ luminosity, which is not seen in the simple dust density model. We therefore suggest that this behavior is not caused solely by variations in dust density but rather is due to a combination of density variations and geometric effects, but a full investigation of this will depend on the availability of inclination-independent measurements of galaxy structure. This is a study we hope to undertake in the future.

\section*{Acknowledgements}

We wish to thank Roelof de Jong and Ned Taylor for valuable discussions, and Dustin Lang for making the \emph{WISE} magnitudes of \citet{lang_etal_14} available for testing before their publication. We also wish to thank the anonymous referee for their helpful suggestions on presenting some aspects of our results. This work was partially supported by NSF-AST 1514835. This research has made use of NASA's Astrophysics Data System Bibliographic Services. This publication makes use of data products from the Wide-field Infrared Survey Explorer, which is a joint project of the University of California, Los Angeles, and the Jet Propulsion Laboratory/California Institute of Technology, funded by the National Aeronautics and Space Administration. This publication makes use of data products from the Two Micron All Sky Survey, which is a joint project of the University of Massachusetts and the Infrared Processing and Analysis Center/California Institute of Technology, funded by the National Aeronautics and Space Administration and the National Science Foundation. Funding for SDSS-III has been provided by the Alfred P. Sloan Foundation, the Participating Institutions, the National Science Foundation, and the U.S. Department of Energy Office of Science. The SDSS-III web site is http://www.sdss3.org/. SDSS-III is managed by the Astrophysical Research Consortium for the Participating Institutions of the SDSS-III Collaboration including the University of Arizona, the Brazilian Participation Group, Brookhaven National Laboratory, Carnegie Mellon University, University of Florida, the French Participation Group, the German Participation Group, Harvard University, the Instituto de Astrofisica de Canarias, the Michigan State/Notre Dame/JINA Participation Group, Johns Hopkins University, Lawrence Berkeley National Laboratory, Max Planck Institute for Astrophysics, Max Planck Institute for Extraterrestrial Physics, New Mexico State University, New York University, Ohio State University, Pennsylvania State University, University of Portsmouth, Princeton University, the Spanish Participation Group, University of Tokyo, University of Utah, Vanderbilt University, University of Virginia, University of Washington, and Yale University. 




\bibliography{main}

\begin{thebibliography}{}
\makeatletter
\relax
\def\mn@urlcharsother{\let\do\@makeother \do\$\do\&\do\#\do\^\do\_\do\%\do\~}
\def\mn@doi{\begingroup\mn@urlcharsother \@ifnextchar [ {\mn@doi@}
  {\mn@doi@[]}}
\def\mn@doi@[#1]#2{\def\@tempa{#1}\ifx\@tempa\@empty \href
  {http://dx.doi.org/#2} {doi:#2}\else \href {http://dx.doi.org/#2} {#1}\fi
  \endgroup}
\def\mn@eprint#1#2{\mn@eprint@#1:#2::\@nil}
\def\mn@eprint@arXiv#1{\href {http://arxiv.org/abs/#1} {{\tt arXiv:#1}}}
\def\mn@eprint@dblp#1{\href {http://dblp.uni-trier.de/rec/bibtex/#1.xml}
  {dblp:#1}}
\def\mn@eprint@#1:#2:#3:#4\@nil{\def\@tempa {#1}\def\@tempb {#2}\def\@tempc
  {#3}\ifx \@tempc \@empty \let \@tempc \@tempb \let \@tempb \@tempa \fi \ifx
  \@tempb \@empty \def\@tempb {arXiv}\fi \@ifundefined
  {mn@eprint@\@tempb}{\@tempb:\@tempc}{\expandafter \expandafter \csname
  mn@eprint@\@tempb\endcsname \expandafter{\@tempc}}}

\bibitem[\protect\citeauthoryear{{Ahn} et~al.,}{{Ahn}
  et~al.}{2014}]{ahn_etal_14}
{Ahn} C.~P.,  et~al., 2014, \mn@doi [\apjs] {10.1088/0067-0049/211/2/17}, \href
  {http://adsabs.harvard.edu/abs/2014ApJS..211...17A} {211, 17}

\bibitem[\protect\citeauthoryear{{Bell}}{{Bell}}{2003}]{bell_03}
{Bell} E.~F.,  2003, \mn@doi [\apj] {10.1086/367829}, \href
  {http://adsabs.harvard.edu/abs/2003ApJ...586..794B} {586, 794}

\bibitem[\protect\citeauthoryear{{Bell} \& {de Jong}}{{Bell} \& {de
  Jong}}{2001}]{bell_dejong_01}
{Bell} E.~F.,  {de Jong} R.~S.,  2001, \mn@doi [\apj] {10.1086/319728}, \href
  {http://adsabs.harvard.edu/abs/2001ApJ...550..212B} {550, 212}

\bibitem[\protect\citeauthoryear{{Berlind}, {Quillen}, {Pogge}  \&
  {Sellgren}}{{Berlind} et~al.}{1997}]{berlind_etal_97}
{Berlind} A.~A.,  {Quillen} A.~C.,  {Pogge} R.~W.,   {Sellgren} K.,  1997,
  \mn@doi [\aj] {10.1086/118457}, \href
  {http://adsabs.harvard.edu/abs/1997AJ....114..107B} {114, 107}

\bibitem[\protect\citeauthoryear{{Blanton} \& {Roweis}}{{Blanton} \&
  {Roweis}}{2007}]{blanton_roweis_07}
{Blanton} M.~R.,  {Roweis} S.,  2007, \mn@doi [\aj] {10.1086/510127}, \href
  {http://adsabs.harvard.edu/abs/2007AJ....133..734B} {133, 734}

\bibitem[\protect\citeauthoryear{{Blanton} et~al.,}{{Blanton}
  et~al.}{2005}]{blanton_etal_05}
{Blanton} M.~R.,  et~al., 2005, \mn@doi [\aj] {10.1086/429803}, \href
  {http://adsabs.harvard.edu/abs/2005AJ....129.2562B} {129, 2562}

\bibitem[\protect\citeauthoryear{{Bruzual A.}, {Magris}  \& {Calvet}}{{Bruzual
  A.} et~al.}{1988}]{bruzual_etal_88}
{Bruzual A.} G.,  {Magris} G.,   {Calvet} N.,  1988, \mn@doi [\apj]
  {10.1086/166776}, \href {http://adsabs.harvard.edu/abs/1988ApJ...333..673B}
  {333, 673}

\bibitem[\protect\citeauthoryear{{Byun}, {Freeman}  \& {Kylafis}}{{Byun}
  et~al.}{1994}]{byun_freeman_kylafis_94}
{Byun} Y.~I.,  {Freeman} K.~C.,   {Kylafis} N.~D.,  1994, \mn@doi [\apj]
  {10.1086/174553}, \href {http://adsabs.harvard.edu/abs/1994ApJ...432..114B}
  {432, 114}

\bibitem[\protect\citeauthoryear{{Calzetti}}{{Calzetti}}{2001}]{calzetti_01}
{Calzetti} D.,  2001, \mn@doi [\nar] {10.1016/S1387-6473(01)00144-0}, \href
  {http://adsabs.harvard.edu/abs/2001NewAR..45..601C} {45, 601}

\bibitem[\protect\citeauthoryear{{Calzetti}}{{Calzetti}}{2013}]{calzetti_13}
{Calzetti} D.,  2013, {Star Formation Rate Indicators}.
p.~419

\bibitem[\protect\citeauthoryear{{Calzetti}, {Armus}, {Bohlin}, {Kinney},
  {Koornneef}  \& {Storchi-Bergmann}}{{Calzetti} et~al.}{2000}]{calzetti_00}
{Calzetti} D.,  {Armus} L.,  {Bohlin} R.~C.,  {Kinney} A.~L.,  {Koornneef} J.,
   {Storchi-Bergmann} T.,  2000, \mn@doi [\apj] {10.1086/308692}, \href
  {http://adsabs.harvard.edu/abs/2000ApJ...533..682C} {533, 682}

\bibitem[\protect\citeauthoryear{{Chang}, {van der Wel}, {da Cunha}  \&
  {Rix}}{{Chang} et~al.}{2015}]{chang_etal_15}
{Chang} Y.-Y.,  {van der Wel} A.,  {da Cunha} E.,   {Rix} H.-W.,  2015, \mn@doi
  [\apjs] {10.1088/0067-0049/219/1/8}, \href
  {http://adsabs.harvard.edu/abs/2015ApJS..219....8C} {219, 8}

\bibitem[\protect\citeauthoryear{{Charlot} \& {Fall}}{{Charlot} \&
  {Fall}}{2000}]{charlot_fall_00}
{Charlot} S.,  {Fall} S.~M.,  2000, \mn@doi [\apj] {10.1086/309250}, \href
  {http://adsabs.harvard.edu/abs/2000ApJ...539..718C} {539, 718}

\bibitem[\protect\citeauthoryear{{Cho} \& {Park}}{{Cho} \&
  {Park}}{2009}]{cho_park_09}
{Cho} J.,  {Park} C.,  2009, \mn@doi [\apj] {10.1088/0004-637X/693/2/1045},
  \href {http://adsabs.harvard.edu/abs/2009ApJ...693.1045C} {693, 1045}

\bibitem[\protect\citeauthoryear{{Cluver} et~al.,}{{Cluver}
  et~al.}{2014}]{cluver_etal_14}
{Cluver} M.~E.,  et~al., 2014, \mn@doi [\apj] {10.1088/0004-637X/782/2/90},
  \href {http://adsabs.harvard.edu/abs/2014ApJ...782...90C} {782, 90}

\bibitem[\protect\citeauthoryear{{Cutri} et~al.,}{{Cutri}
  et~al.}{2006}]{cutri_etal_06}
{Cutri} R.~M.,  et~al., 2006, Explanatory Supplement to the 2MASS All Sky Data
  Release and Extended Mission Products, \url
  {http://www.ipac.caltech.edu/2mass/releases/allsky/doc/explsup.html}

\bibitem[\protect\citeauthoryear{{Disney}, {Davies}  \& {Phillipps}}{{Disney}
  et~al.}{1989}]{disney_davies_phillipps_89}
{Disney} M.,  {Davies} J.,   {Phillipps} S.,  1989, \mnras, \href
  {http://adsabs.harvard.edu/abs/1989MNRAS.239..939D} {239, 939}

\bibitem[\protect\citeauthoryear{{Draine} \& {Li}}{{Draine} \&
  {Li}}{2007}]{draine_li_07}
{Draine} B.~T.,  {Li} A.,  2007, \mn@doi [\apj] {10.1086/511055}, \href
  {http://adsabs.harvard.edu/abs/2007ApJ...657..810D} {657, 810}

\bibitem[\protect\citeauthoryear{{Draine} et~al.,}{{Draine}
  et~al.}{2007}]{draine_etal_07}
{Draine} B.~T.,  et~al., 2007, \mn@doi [\apj] {10.1086/518306}, \href
  {http://adsabs.harvard.edu/abs/2007ApJ...663..866D} {663, 866}

\bibitem[\protect\citeauthoryear{{Driver}, {Popescu}, {Tuffs}, {Liske},
  {Graham}, {Allen}  \& {de Propris}}{{Driver} et~al.}{2007}]{driver_etal_07}
{Driver} S.~P.,  {Popescu} C.~C.,  {Tuffs} R.~J.,  {Liske} J.,  {Graham} A.~W.,
   {Allen} P.~D.,   {de Propris} R.,  2007, \mn@doi [\mnras]
  {10.1111/j.1365-2966.2007.11862.x}, \href
  {http://adsabs.harvard.edu/abs/2007MNRAS.379.1022D} {379, 1022}

\bibitem[\protect\citeauthoryear{{Driver} et~al.,}{{Driver}
  et~al.}{2012}]{driver_etal_12}
{Driver} S.~P.,  et~al., 2012, \mn@doi [\mnras]
  {10.1111/j.1365-2966.2012.22036.x}, \href
  {http://adsabs.harvard.edu/abs/2012MNRAS.427.3244D} {427, 3244}

\bibitem[\protect\citeauthoryear{{Eisenstein} et~al.,}{{Eisenstein}
  et~al.}{2011}]{eisenstein_etal_11}
{Eisenstein} D.~J.,  et~al., 2011, \mn@doi [\aj] {10.1088/0004-6256/142/3/72},
  \href {http://adsabs.harvard.edu/abs/2011AJ....142...72E} {142, 72}

\bibitem[\protect\citeauthoryear{{Emsellem} et~al.,}{{Emsellem}
  et~al.}{2011}]{emsellem_etal_11}
{Emsellem} E.,  et~al., 2011, \mn@doi [\mnras]
  {10.1111/j.1365-2966.2011.18496.x}, \href
  {http://adsabs.harvard.edu/abs/2011MNRAS.414..888E} {414, 888}

\bibitem[\protect\citeauthoryear{{Fontanot}, {Somerville}, {Silva}, {Monaco}
  \& {Skibba}}{{Fontanot} et~al.}{2009}]{fontanot_etal_09}
{Fontanot} F.,  {Somerville} R.~S.,  {Silva} L.,  {Monaco} P.,   {Skibba} R.,
  2009, \mn@doi [\mnras] {10.1111/j.1365-2966.2008.14126.x}, \href
  {http://adsabs.harvard.edu/abs/2009MNRAS.392..553F} {392, 553}

\bibitem[\protect\citeauthoryear{{Giovanelli}, {Haynes}, {Salzer}, {Wegner},
  {da Costa}  \& {Freudling}}{{Giovanelli} et~al.}{1994}]{giovanelli_etal_94}
{Giovanelli} R.,  {Haynes} M.~P.,  {Salzer} J.~J.,  {Wegner} G.,  {da Costa}
  L.~N.,   {Freudling} W.,  1994, \mn@doi [\aj] {10.1086/117014}, \href
  {http://adsabs.harvard.edu/abs/1994AJ....107.2036G} {107, 2036}

\bibitem[\protect\citeauthoryear{{Giovanelli}, {Haynes}, {Salzer}, {Wegner},
  {da Costa}  \& {Freudling}}{{Giovanelli} et~al.}{1995}]{giovanelli_etal_95}
{Giovanelli} R.,  {Haynes} M.~P.,  {Salzer} J.~J.,  {Wegner} G.,  {da Costa}
  L.~N.,   {Freudling} W.,  1995, \mn@doi [\aj] {10.1086/117586}, \href
  {http://adsabs.harvard.edu/abs/1995AJ....110.1059G} {110, 1059}

\bibitem[\protect\citeauthoryear{{Holwerda}, {Gonzalez}, {Allen}  \& {van der
  Kruit}}{{Holwerda} et~al.}{2005}]{holwerda_etal_05}
{Holwerda} B.~W.,  {Gonzalez} R.~A.,  {Allen} R.~J.,   {van der Kruit} P.~C.,
  2005, \mn@doi [\aj] {10.1086/427716}, \href
  {http://adsabs.harvard.edu/abs/2005AJ....129.1396H} {129, 1396}

\bibitem[\protect\citeauthoryear{{Holwerda} et~al.,}{{Holwerda}
  et~al.}{2007}]{holwerda_etal_07}
{Holwerda} B.~W.,  et~al., 2007, \mn@doi [\aj] {10.1086/522230}, \href
  {http://adsabs.harvard.edu/abs/2007AJ....134.2226H} {134, 2226}

\bibitem[\protect\citeauthoryear{{Huizinga} \& {van Albada}}{{Huizinga} \& {van
  Albada}}{1992}]{huizinga_vanalbada_92}
{Huizinga} J.~E.,  {van Albada} T.~S.,  1992, \mnras, \href
  {http://adsabs.harvard.edu/abs/1992MNRAS.254..677H} {254, 677}

\bibitem[\protect\citeauthoryear{{Jonsson}, {Cox}, {Primack}  \&
  {Somerville}}{{Jonsson} et~al.}{2006}]{jonsson_etal_06}
{Jonsson} P.,  {Cox} T.~J.,  {Primack} J.~R.,   {Somerville} R.~S.,  2006,
  \mn@doi [\apj] {10.1086/497567}, \href
  {http://adsabs.harvard.edu/abs/2006ApJ...637..255J} {637, 255}

\bibitem[\protect\citeauthoryear{{Keel}, {Manning}, {Holwerda}, {Lintott}  \&
  {Schawinski}}{{Keel} et~al.}{2014}]{keel_etal_14}
{Keel} W.~C.,  {Manning} A.~M.,  {Holwerda} B.~W.,  {Lintott} C.~J.,
  {Schawinski} K.,  2014, \mn@doi [\aj] {10.1088/0004-6256/147/2/44}, \href
  {http://adsabs.harvard.edu/abs/2014AJ....147...44K} {147, 44}

\bibitem[\protect\citeauthoryear{{Kennicutt}}{{Kennicutt}}{1989}]{kennicutt_89}
{Kennicutt} Jr. R.~C.,  1989, \mn@doi [\apj] {10.1086/167834}, \href
  {http://adsabs.harvard.edu/abs/1989ApJ...344..685K} {344, 685}

\bibitem[\protect\citeauthoryear{{Kennicutt}}{{Kennicutt}}{1998}]{kennicutt_98}
{Kennicutt} Jr. R.~C.,  1998, \mn@doi [\apj] {10.1086/305588}, \href
  {http://adsabs.harvard.edu/abs/1998ApJ...498..541K} {498, 541}

\bibitem[\protect\citeauthoryear{{Lang}, {Hogg}  \& {Schlegel}}{{Lang}
  et~al.}{2014}]{lang_etal_14}
{Lang} D.,  {Hogg} D.~W.,   {Schlegel} D.~J.,  2014, \aj, \href
  {http://adsabs.harvard.edu/abs/2014arXiv1410.7397L} {Submitted}

\bibitem[\protect\citeauthoryear{{Lawrence} et~al.,}{{Lawrence}
  et~al.}{2007}]{lawrence_etal_07}
{Lawrence} A.,  et~al., 2007, \mn@doi [\mnras]
  {10.1111/j.1365-2966.2007.12040.x}, \href
  {http://adsabs.harvard.edu/abs/2007MNRAS.379.1599L} {379, 1599}

\bibitem[\protect\citeauthoryear{{Li} \& {Draine}}{{Li} \&
  {Draine}}{2001}]{li_draine_01}
{Li} A.,  {Draine} B.~T.,  2001, \mn@doi [\apj] {10.1086/323147}, \href
  {http://adsabs.harvard.edu/abs/2001ApJ...554..778L} {554, 778}

\bibitem[\protect\citeauthoryear{{Liu} et~al.,}{{Liu}
  et~al.}{2013}]{liu_etal_13}
{Liu} G.,  et~al., 2013, \mn@doi [\apjl] {10.1088/2041-8205/778/2/L41}, \href
  {http://adsabs.harvard.edu/abs/2013ApJ...778L..41L} {778, L41}

\bibitem[\protect\citeauthoryear{{Maller}, {Berlind}, {Blanton}  \&
  {Hogg}}{{Maller} et~al.}{2009}]{maller_etal_09}
{Maller} A.~H.,  {Berlind} A.~A.,  {Blanton} M.~R.,   {Hogg} D.~W.,  2009,
  \mn@doi [\apj] {10.1088/0004-637X/691/1/394}, \href
  {http://adsabs.harvard.edu/abs/2009ApJ...691..394M} {691, 394}

\bibitem[\protect\citeauthoryear{{Masters} et~al.,}{{Masters}
  et~al.}{2010}]{masters_etal_10}
{Masters} K.~L.,  et~al., 2010, \mn@doi [\mnras]
  {10.1111/j.1365-2966.2010.16335.x}, \href
  {http://adsabs.harvard.edu/abs/2010MNRAS.404..792M} {404, 792}

\bibitem[\protect\citeauthoryear{{Meidt} et~al.,}{{Meidt}
  et~al.}{2012}]{meidt_etal_12}
{Meidt} S.~E.,  et~al., 2012, \mn@doi [\apj] {10.1088/0004-637X/744/1/17},
  \href {http://adsabs.harvard.edu/abs/2012ApJ...744...17M} {744, 17}

\bibitem[\protect\citeauthoryear{{Meidt} et~al.,}{{Meidt}
  et~al.}{2014}]{meidt_etal_14}
{Meidt} S.~E.,  et~al., 2014, \mn@doi [\apj] {10.1088/0004-637X/788/2/144},
  \href {http://adsabs.harvard.edu/abs/2014ApJ...788..144M} {788, 144}

\bibitem[\protect\citeauthoryear{{M{\'e}nard}, {Scranton}, {Fukugita}  \&
  {Richards}}{{M{\'e}nard} et~al.}{2010}]{menard_etal_10}
{M{\'e}nard} B.,  {Scranton} R.,  {Fukugita} M.,   {Richards} G.,  2010,
  \mn@doi [\mnras] {10.1111/j.1365-2966.2010.16486.x}, \href
  {http://adsabs.harvard.edu/abs/2010MNRAS.405.1025M} {405, 1025}

\bibitem[\protect\citeauthoryear{{Noeske} et~al.,}{{Noeske}
  et~al.}{2007}]{noeske_etal_07}
{Noeske} K.~G.,  et~al., 2007, \mn@doi [\apjl] {10.1086/517926}, \href
  {http://adsabs.harvard.edu/abs/2007ApJ...660L..43N} {660, L43}

\bibitem[\protect\citeauthoryear{{Oke} \& {Gunn}}{{Oke} \&
  {Gunn}}{1983}]{oke_gunn_83}
{Oke} J.~B.,  {Gunn} J.~E.,  1983, \mn@doi [\apj] {10.1086/160817}, \href
  {http://adsabs.harvard.edu/abs/1983ApJ...266..713O} {266, 713}

\bibitem[\protect\citeauthoryear{{Papovich} \& {Bell}}{{Papovich} \&
  {Bell}}{2002}]{papovich_bell_02}
{Papovich} C.,  {Bell} E.~F.,  2002, \mn@doi [\apjl] {10.1086/344814}, \href
  {http://adsabs.harvard.edu/abs/2002ApJ...579L...1P} {579, L1}

\bibitem[\protect\citeauthoryear{{Pastrav}, {Popescu}, {Tuffs}  \&
  {Sansom}}{{Pastrav} et~al.}{2013}]{pastrav_etal_13}
{Pastrav} B.~A.,  {Popescu} C.~C.,  {Tuffs} R.~J.,   {Sansom} A.~E.,  2013,
  \mn@doi [\aap] {10.1051/0004-6361/201220962}, \href
  {http://adsabs.harvard.edu/abs/2013A%26A...553A..80P} {553, A80}

\bibitem[\protect\citeauthoryear{{Popescu}, {Misiriotis}, {Kylafis}, {Tuffs}
  \& {Fischera}}{{Popescu} et~al.}{2000}]{popescu_etal_00}
{Popescu} C.~C.,  {Misiriotis} A.,  {Kylafis} N.~D.,  {Tuffs} R.~J.,
  {Fischera} J.,  2000, \aap, \href
  {http://adsabs.harvard.edu/abs/2000A%26A...362..138P} {362, 138}

\bibitem[\protect\citeauthoryear{{Salim} et~al.,}{{Salim}
  et~al.}{2007}]{salim_etal_07}
{Salim} S.,  et~al., 2007, \mn@doi [\apjs] {10.1086/519218}, \href
  {http://adsabs.harvard.edu/abs/2007ApJS..173..267S} {173, 267}

\bibitem[\protect\citeauthoryear{{Schechtman-Rook}, {Bershady}  \&
  {Wood}}{{Schechtman-Rook} et~al.}{2012}]{schechtman-rook_etal_12}
{Schechtman-Rook} A.,  {Bershady} M.~A.,   {Wood} K.,  2012, \mn@doi [\apj]
  {10.1088/0004-637X/746/1/70}, \href
  {http://adsabs.harvard.edu/abs/2012ApJ...746...70S} {746, 70}

\bibitem[\protect\citeauthoryear{{S{\'e}rsic}}{{S{\'e}rsic}}{1963}]{sersic_63}
{S{\'e}rsic} J.~L.,  1963, Boletin de la Asociacion Argentina de Astronomia La
  Plata Argentina, \href {http://adsabs.harvard.edu/abs/1963BAAA....6...41S}
  {6, 41}

\bibitem[\protect\citeauthoryear{{Simard}, {Mendel}, {Patton}, {Ellison}  \&
  {McConnachie}}{{Simard} et~al.}{2011}]{simard_etal_11}
{Simard} L.,  {Mendel} J.~T.,  {Patton} D.~R.,  {Ellison} S.~L.,
  {McConnachie} A.~W.,  2011, \mn@doi [\apjs] {10.1088/0067-0049/196/1/11},
  \href {http://adsabs.harvard.edu/abs/2011ApJS..196...11S} {196, 11}

\bibitem[\protect\citeauthoryear{{Skrutskie} et~al.,}{{Skrutskie}
  et~al.}{2006}]{skrutskie_etal_06}
{Skrutskie} M.~F.,  et~al., 2006, \mn@doi [\aj] {10.1086/498708}, \href
  {http://adsabs.harvard.edu/abs/2006AJ....131.1163S} {131, 1163}

\bibitem[\protect\citeauthoryear{{Soifer} \& {Neugebauer}}{{Soifer} \&
  {Neugebauer}}{1991}]{soifer_neugebauer_91}
{Soifer} B.~T.,  {Neugebauer} G.,  1991, \mn@doi [\aj] {10.1086/115691}, \href
  {http://adsabs.harvard.edu/abs/1991AJ....101..354S} {101, 354}

\bibitem[\protect\citeauthoryear{{Steinacker}, {Baes}  \&
  {Gordon}}{{Steinacker} et~al.}{2013}]{steinacker_etal_13}
{Steinacker} J.,  {Baes} M.,   {Gordon} K.~D.,  2013, \mn@doi [\araa]
  {10.1146/annurev-astro-082812-141042}, \href
  {http://adsabs.harvard.edu/abs/2013ARA%26A..51...63S} {51, 63}

\bibitem[\protect\citeauthoryear{{Strauss} et~al.,}{{Strauss}
  et~al.}{2002}]{strauss_etal_02}
{Strauss} M.~A.,  et~al., 2002, \mn@doi [\aj] {10.1086/342343}, \href
  {http://adsabs.harvard.edu/abs/2002AJ....124.1810S} {124, 1810}

\bibitem[\protect\citeauthoryear{{Tremblay} \& {Merritt}}{{Tremblay} \&
  {Merritt}}{1996}]{tremblay_merritt_96}
{Tremblay} B.,  {Merritt} D.,  1996, \mn@doi [\aj] {10.1086/117959}, \href
  {http://adsabs.harvard.edu/abs/1996AJ....111.2243T} {111, 2243}

\bibitem[\protect\citeauthoryear{{Tremonti} et~al.,}{{Tremonti}
  et~al.}{2004}]{tremonti_etal_04}
{Tremonti} C.~A.,  et~al., 2004, \mn@doi [\apj] {10.1086/423264}, \href
  {http://adsabs.harvard.edu/abs/2004ApJ...613..898T} {613, 898}

\bibitem[\protect\citeauthoryear{{Tuffs}, {Popescu}, {V{\"o}lk}, {Kylafis}  \&
  {Dopita}}{{Tuffs} et~al.}{2004}]{tuffs_etal_04}
{Tuffs} R.~J.,  {Popescu} C.~C.,  {V{\"o}lk} H.~J.,  {Kylafis} N.~D.,
  {Dopita} M.~A.,  2004, \mn@doi [\aap] {10.1051/0004-6361:20035689}, \href
  {http://adsabs.harvard.edu/abs/2004A%26A...419..821T} {419, 821}

\bibitem[\protect\citeauthoryear{{Tully}, {Pierce}, {Huang}, {Saunders},
  {Verheijen}  \& {Witchalls}}{{Tully} et~al.}{1998}]{tully_etal_98}
{Tully} R.~B.,  {Pierce} M.~J.,  {Huang} J.-S.,  {Saunders} W.,  {Verheijen}
  M.~A.~W.,   {Witchalls} P.~L.,  1998, \mn@doi [\aj] {10.1086/300379}, \href
  {http://adsabs.harvard.edu/abs/1998AJ....115.2264T} {115, 2264}

\bibitem[\protect\citeauthoryear{{Valentijn}}{{Valentijn}}{1990}]{valentijn_90}
{Valentijn} E.~A.,  1990, \mn@doi [\nat] {10.1038/346153a0}, \href
  {http://adsabs.harvard.edu/abs/1990Natur.346..153V} {346, 153}

\bibitem[\protect\citeauthoryear{{Vincent} \& {Ryden}}{{Vincent} \&
  {Ryden}}{2005}]{vincent_ryden_05}
{Vincent} R.~A.,  {Ryden} B.~S.,  2005, \mn@doi [\apj] {10.1086/428765}, \href
  {http://adsabs.harvard.edu/abs/2005ApJ...623..137V} {623, 137}

\bibitem[\protect\citeauthoryear{{Wen}, {Wu}, {Zhu}, {Lam}, {Wu}, {Wicker},
  {Long}  \& {Zhao}}{{Wen} et~al.}{2014}]{wen_etal_14}
{Wen} X.-Q.,  {Wu} H.,  {Zhu} Y.-N.,  {Lam} M.~I.,  {Wu} C.-J.,  {Wicker} J.,
  {Long} R.~J.,   {Zhao} Y.-H.,  2014, \mn@doi [\mnras]
  {10.1093/mnras/stt2112}, \href
  {http://adsabs.harvard.edu/abs/2014MNRAS.438...97W} {438, 97}

\bibitem[\protect\citeauthoryear{{White}, {Keel}  \& {Conselice}}{{White}
  et~al.}{2000}]{white_keel_conselice_00}
{White} III R.~E.,  {Keel} W.~C.,   {Conselice} C.~J.,  2000, \mn@doi [\apj]
  {10.1086/317011}, \href {http://adsabs.harvard.edu/abs/2000ApJ...542..761W}
  {542, 761}

\bibitem[\protect\citeauthoryear{{Wild}, {Charlot}, {Brinchmann}, {Heckman},
  {Vince}, {Pacifici}  \& {Chevallard}}{{Wild} et~al.}{2011}]{wild_etal_11}
{Wild} V.,  {Charlot} S.,  {Brinchmann} J.,  {Heckman} T.,  {Vince} O.,
  {Pacifici} C.,   {Chevallard} J.,  2011, \mn@doi [\mnras]
  {10.1111/j.1365-2966.2011.19367.x}, \href
  {http://adsabs.harvard.edu/abs/2011MNRAS.417.1760W} {417, 1760}

\bibitem[\protect\citeauthoryear{{Witt} \& {Gordon}}{{Witt} \&
  {Gordon}}{2000}]{witt_gordon_00}
{Witt} A.~N.,  {Gordon} K.~D.,  2000, \mn@doi [\apj] {10.1086/308197}, \href
  {http://adsabs.harvard.edu/abs/2000ApJ...528..799W} {528, 799}

\bibitem[\protect\citeauthoryear{{Wright} et~al.,}{{Wright}
  et~al.}{2010}]{wright_etal_10}
{Wright} E.~L.,  et~al., 2010, \mn@doi [\aj] {10.1088/0004-6256/140/6/1868},
  \href {http://adsabs.harvard.edu/abs/2010AJ....140.1868W} {140, 1868}

\bibitem[\protect\citeauthoryear{{de Jong}}{{de Jong}}{1996}]{dejong_96}
{de Jong} R.~S.,  1996, \aap, \href
  {http://adsabs.harvard.edu/abs/1996A%26A...313..377D} {313, 377}

\bibitem[\protect\citeauthoryear{{de Jong} \& {Lacey}}{{de Jong} \&
  {Lacey}}{2000}]{dejong_lacey_00}
{de Jong} R.~S.,  {Lacey} C.,  2000, \mn@doi [\apj] {10.1086/317840}, \href
  {http://adsabs.harvard.edu/abs/2000ApJ...545..781D} {545, 781}

\bibitem[\protect\citeauthoryear{{van der Wel}, {Rix}, {Holden}, {Bell}  \&
  {Robaina}}{{van der Wel} et~al.}{2009}]{vanderwel_etal_09}
{van der Wel} A.,  {Rix} H.-W.,  {Holden} B.~P.,  {Bell} E.~F.,   {Robaina}
  A.~R.,  2009, \mn@doi [\apjl] {10.1088/0004-637X/706/1/L120}, \href
  {http://adsabs.harvard.edu/abs/2009ApJ...706L.120V} {706, L120}

\makeatother
\end{thebibliography}



\appendix
\section{Attenuation data tables}

\begin{table*}
\caption{Attenuation parameter $\gamma$ measurements and uncertainties, u band.}
\label{tab:atten_u}
\begin{tabular}{dcpppppp}
\hline
 & & \multicolumn{6}{c}{$M_{3.4\um}$} \\
{[}12]=[3.4] & & \multicolumn{1}{d}{(-24)=(-23)} & \multicolumn{1}{d}{(-23)=(-22)} & \multicolumn{1}{d}{(-22)=(-21)} & \multicolumn{1}{d}{(-21)=(-20)} & \multicolumn{1}{d}{(-20)=(-19)} & \multicolumn{1}{d}{(-19)=(-18)}\\
\hline
(3.0)=(3.5)   & & & -0.72+0.37 & & & & \\
(2.5)=(3.0)   & & -0.64+0.59 & -0.15+0.16 & 0.123+0.068 & 0.066+0.073 & 0.13+0.12 & \\
(2.0)=(2.5)   & & -0.38+0.24 & -0.003+0.071 & 0.087+0.054 & 0.008+0.057 & -0.032+0.074 & \\ 
(1.5)=(2.0)   & & -0.78+0.47 & -0.071+0.078 & 0.223+0.060 & 0.138+0.069 & 0.00+0.12 & \\
(1.0)=(1.5)   & & & 0.113+0.087 & 0.230+0.064 & 0.389+0.085 & & \\
(0.5)=(1.0)   & & & 0.452+0.092 & 0.467+0.071 & 0.532+0.081 & 0.80+0.12 & \\
(0.0)=(0.5)   & & & 0.745+0.080 & 0.926+0.066 & 1.007+0.076 & 0.79+0.19 & \\
(-0.5)=(0.0) & & & 0.973+0.086 & 1.168+0.063 & 1.236+0.069 & 1.060+0.090 & 0.64+0.15 \\
(-1.0)=(-0.5) & & & 1.385+0.095 & 1.574+0.067 & 1.647+0.067 & 1.42+0.11 & 1.15+0.33 \\
(-1.5)=(-1.0) & & & 1.46+0.18 & 1.81+0.13 & 1.72+0.11 & 1.33+0.23 & \\
\hline
\end{tabular}
\end{table*}

\begin{table*}
\caption{Attenuation parameter $\gamma$ measurements and uncertainties, g band.}
\label{tab:atten_g}
\begin{tabular}{dcpppppp}
\hline
 & & \multicolumn{6}{c}{$M_{3.4\um}$} \\
{[}12]=[3.4] & & \multicolumn{1}{d}{(-24)=(-23)} & \multicolumn{1}{d}{(-23)=(-22)} & \multicolumn{1}{d}{(-22)=(-21)} & \multicolumn{1}{d}{(-21)=(-20)} & \multicolumn{1}{d}{(-20)=(-19)} & \multicolumn{1}{d}{(-19)=(-18)}\\
\hline
(3.0)=(3.5)   & &            & -0.38+0.17 & & & & \\
(2.5)=(3.0)   & & -0.41+0.23 & -0.015+0.088 & 0.131+0.056 & 0.108+0.059 & 0.167+0.087 & \\
(2.0)=(2.5)   & & -0.30+0.13 & 0.048+0.057 & 0.137+0.052 & 0.060+0.053 & 0.036+0.065 & \\ 
(1.5)=(2.0)   & & -0.22+0.22 & 0.119+0.061 & 0.200+0.054 & 0.176+0.059 & 0.076+0.082 & \\
(1.0)=(1.5)   & &            & 0.275+0.063 & 0.298+0.056 & 0.398+0.065 & & \\
(0.5)=(1.0)   & &            & 0.457+0.063 & 0.457+0.058 & 0.502+0.062 & 0.662+0.083 & \\
(0.0)=(0.5)   & &            & 0.630+0.063 & 0.775+0.057 & 0.790+0.062 & 0.62+0.12 & \\
(-0.5)=(0.0)  & &            & 0.762+0.063 & 0.919+0.055 & 0.974+0.059 & 0.890+0.071 & 0.58+0.12 \\
(-1.0)=(-0.5) & &            & 1.005+0.070 & 1.144+0.056 & 1.212+0.059 & 1.093+0.081 & 1.06+0.24 \\
(-1.5)=(-1.0) & &            & 1.04+0.13 & 1.244+0.088 & 1.238+0.086 & 0.93+0.17 & \\
\hline
\end{tabular}
\end{table*}

\begin{table*}
\caption{Attenuation parameter $\gamma$ measurements and uncertainties, r band.}
\label{tab:atten_r}
\begin{tabular}{dcpppppp}
\hline
 & & \multicolumn{6}{c}{$M_{3.4\um}$} \\
{[}12]=[3.4] & & \multicolumn{1}{d}{(-24)=(-23)} & \multicolumn{1}{d}{(-23)=(-22)} & \multicolumn{1}{d}{(-22)=(-21)} & \multicolumn{1}{d}{(-21)=(-20)} & \multicolumn{1}{d}{(-20)=(-19)} & \multicolumn{1}{d}{(-19)=(-18)}\\
\hline
(3.0)=(3.5)   & &            & -0.347+0.17 & & & & \\
(2.5)=(3.0)   & & -0.66+0.24 & 0.026+0.03 & 0.131+0.055 & 0.098+0.057 & 0.136+0.080 & \\
(2.0)=(2.5)   & & -0.32+0.14 & 0.054+0.056 & 0.141+0.052 & 0.072+0.052 & 0.035+0.062 & \\ 
(1.5)=(2.0)   & & -0.22+0.23 & 0.143+0.059 & 0.178+0.053 & 0.162+0.056 & 0.083+0.075 & \\
(1.0)=(1.5)   & &            & 0.281+0.061 & 0.279+0.055 & 0.344+0.061 & & \\
(0.5)=(1.0)   & &            & 0.438+0.058 & 0.423+0.056 & 0.439+0.058 & 0.577+0.075 & \\
(0.0)=(0.5)   & &            & 0.583+0.059 & 0.689+0.054 & 0.697+0.058 & 0.553+0.096 & \\
(-0.5)=(0.0)  & &            & 0.713+0.060 & 0.824+0.054 & 0.853+0.056 & 0.781+0.065 & 0.53+0.10 \\
(-1.0)=(-0.5) & &            & 0.897+0.064 & 0.982+0.054 & 1.023+0.056 & 0.908+0.073 & 0.91+0.19 \\
(-1.5)=(-1.0) & &            & 0.87+0.11 & 1.002+0.076 & 0.982+0.076 & 0.80+0.14 & \\
\hline
\end{tabular}
\end{table*}

\begin{table*}
\caption{Attenuation parameter $\gamma$ measurements and uncertainties, i band.}
\label{tab:atten_i}
\begin{tabular}{dcpppppp}
\hline
 & & \multicolumn{6}{c}{$M_{3.4\um}$} \\
{[}12]=[3.4] & & \multicolumn{1}{d}{(-24)=(-23)} & \multicolumn{1}{d}{(-23)=(-22)} & \multicolumn{1}{d}{(-22)=(-21)} & \multicolumn{1}{d}{(-21)=(-20)} & \multicolumn{1}{d}{(-20)=(-19)} & \multicolumn{1}{d}{(-19)=(-18)}\\
\hline
(3.0)=(3.5)   & &            & -0.314+0.17 & & & & \\
(2.5)=(3.0)   & & -0.57+0.22 & 0.024+0.081 & 0.108+0.055 & 0.071+0.056 & 0.114+0.080 & \\
(2.0)=(2.5)   & & -0.33+0.13 & 0.035+0.055 & 0.116+0.051 & 0.050+0.052 & 0.026+0.060 & \\ 
(1.5)=(2.0)   & & -0.24+0.22 & 0.095+0.058 & 0.133+0.053 & 0.127+0.055 & 0.057+0.074 & \\
(1.0)=(1.5)   & &            & 0.210+0.059 & 0.216+0.054 & 0.269+0.060 & & \\
(0.5)=(1.0)   & &            & 0.336+0.057 & 0.343+0.054 & 0.346+0.057 & 0.504+0.073 & \\
(0.0)=(0.5)   & &            & 0.449+0.057 & 0.568+0.053 & 0.581+0.057 & 0.480+0.085 & \\
(-0.5)=(0.0)  & &            & 0.539+0.058 & 0.681+0.053 & 0.736+0.056 & 0.712+0.062 & 0.489+0.097 \\
(-1.0)=(-0.5) & &            & 0.694+0.063 & 0.793+0.054 & 0.872+0.055 & 0.803+0.070 & 0.85+0.18 \\
(-1.5)=(-1.0) & &            & 0.594+0.097 & 0.802+0.072 & 0.804+0.071 & 0.67+0.13 & \\
\hline
\end{tabular}
\end{table*}

\begin{table*}
\caption{Attenuation parameter $\gamma$ measurements and uncertainties, z band.}
\label{tab:atten_z}
\begin{tabular}{dcpppppp}
\hline
 & & \multicolumn{6}{c}{$M_{3.4\um}$} \\
{[}12]=[3.4] & & \multicolumn{1}{d}{(-24)=(-23)} & \multicolumn{1}{d}{(-23)=(-22)} & \multicolumn{1}{d}{(-22)=(-21)} & \multicolumn{1}{d}{(-21)=(-20)} & \multicolumn{1}{d}{(-20)=(-19)} & \multicolumn{1}{d}{(-19)=(-18)}\\
\hline
(3.0)=(3.5)   & &            & -0.31+0.17 & & & & \\
(2.5)=(3.0)   & & -0.60+0.25 & -0.077+0.079 & -0.016+0.055 & -0.040+0.056 & 0.001+0.083 & \\
(2.0)=(2.5)   & & -0.32+0.14 & -0.069+0.056 & -0.016+0.051 & -0.070+0.052 & -0.080+0.060 & \\ 
(1.5)=(2.0)   & & -0.24+0.21 & -0.046+0.058 & -0.022+0.052 & -0.024+0.055 & -0.044+0.072 & \\
(1.0)=(1.5)   & &            & 0.047+0.058 & 0.046+0.054 & 0.087+0.059 & & \\
(0.5)=(1.0)   & &            & 0.154+0.057 & 0.170+0.054 & 0.165+0.057 & 0.329+0.073 & \\
(0.0)=(0.5)   & &            & 0.293+0.057 & 0.378+0.053 & 0.400+0.056 & 0.366+0.080 & \\
(-0.5)=(0.0)  & &            & 0.399+0.059 & 0.539+0.053 & 0.597+0.055 & 0.599+0.061 & 0.398+0.094 \\
(-1.0)=(-0.5) & &            & 0.602+0.062 & 0.663+0.054 & 0.740+0.055 & 0.678+0.066 & 0.71+0.16 \\
(-1.5)=(-1.0) & &            & 0.438+0.098 & 0.674+0.071 & 0.650+0.066 & 0.54+0.12 & \\
\hline
\end{tabular}
\end{table*}

\bsp	
\label{lastpage}
\end{document}